\documentclass[a4paper,11pt]{article}
\pdfoutput=1 

\usepackage{jheppub} 

\usepackage[T1]{fontenc} 

\usepackage{braket} 

\usepackage{mathrsfs}

\newcommand{\tr}{\mathcal{T}}
\newcommand{\sr}{\mathcal{S}}
\newcommand{\de}{\text{d}}
\newcommand{\stl}{r}

\usepackage{tikz}
\usetikzlibrary{math}
\usetikzlibrary{arrows,shapes,positioning}
\usetikzlibrary{decorations.markings}
\tikzstyle arrowstyle=[scale=1]
\tikzstyle directed=[postaction={decorate,decoration={markings,
    mark=at position .65 with {\arrow[arrowstyle]{stealth}}}}]
\tikzstyle reverse directed=[postaction={decorate,decoration={markings,
    mark=at position .65 with {\arrowreversed[arrowstyle]{stealth};}}}]

\newlength{\mywidth}
\usepackage[colorlinks=true
,urlcolor=blue
,anchorcolor=blue
,citecolor=blue
,filecolor=blue
,linkcolor=blue
,menucolor=blue
,linktocpage=true
,pdfproducer=medialab
,pdfa=true
]{hyperref}

\title{\boldmath On the worldsheet S~matrix of the AdS$_3$/CFT$_2$ mixed-flux mirror model}

\author[a]{Nicola Baglioni,}
\author[a,b]{Davide Polvara,}
\author[a]{Andrea Pone,}
\author[a,b]{Alessandro Sfondrini}

\affiliation[a]{Dipartimento di Fisica e Astronomia, Universit\`a degli Studi di Padova,\\
via Marzolo 8, 35131 Padova, Italy}
\affiliation[b]{
Istituto Nazionale di Fisica Nucleare, Sezione di Padova,\\
via Marzolo 8, 35131 Padova, Italy}

\emailAdd{nicola.baglioni@studenti.unipd.it}
\emailAdd{davide.polvara@unipd.it}
\emailAdd{andrea.pone@studenti.unipd.it}
\emailAdd{alessandro.sfondrini@unipd.it}

\abstract{
String on $AdS_3\times S^3\times T^4$ backgrounds are known to be classically integrable in the presence of a mixture of Ramond-Ramond and Neveu-Schwarz-Neveu-Schwarz fluxes. It is expected that this results in the existence of a well-defined factorised worldsheet S~matrix. In order to use integrability to compute the string spectrum we need such a factorised S matrix to exist also for the ``mirror'' model, obtained by a double Wick rotation of the original worldsheet theory.
In the mixed-flux case the mirror model has a complex Hamiltonian, which raises questions on its well-definedness.
In the paper we study the worldsheet tree-level S matrix of the original and mirror model and discuss some necessary conditions for the integrability and reality of the spectrum.
}

\begin{document} 
\maketitle
\flushbottom

\section{Introduction and conclusions}
\label{sec:intro}

Classical integrability for a string non-linear sigma model can be formulated in terms of the existence of a Lax connection~\cite{Bena:2003wd}.
 At the quantum level, integrability can be understood in terms of the factorisation of scattering. In the case at hand, this is the scattering on the two-dimensional worldsheet of the string, once a suitable gauge-fixing is applied~\cite{Arutyunov:2005hd,Arutyunov:2009ga}.
The resulting S~matrix can be compared with perturbative computations made in the near-BMN~\cite{Berenstein:2002jq} kinematics~\cite{Klose:2006zd}, see also~\cite{Arutyunov:2009ga,Beisert:2010jr} for reviews on the integrability approach for worldsheet string theory.

The worldsheet S~matrix is an important ingredient to compute the string spectrum by Bethe ansatz techniques. While the (asymptotic) Bethe-Yang equations capture the large-worldsheet-volume (\textit{i.e.}, large-$R$-charge) spectrum~\cite{Arutyunov:2004vx}, it is important to account for finite-size ``wrapping'' effects~\cite{Ambjorn:2005wa}. At leading order, these are encoded in L\"uscher corrections~\cite{Luscher:1985dn,Luscher:1986pf}, but to obtain the exact spectrum it is necessary to use the thermodynamic Bethe ansatz~\cite{Zamolodchikov:1989cf}, which describes the finite-temperature features of a theory. Strictly speaking, because we are interested in a \textit{finite-volume}, rather than \textit{finite-temperature} theory, we have to perform a double Wick rotation sending the worldsheet temporal and spatial coordinates $(\tau,\sigma)$ to
\begin{equation}
\label{eq:doublewick_tau_sig}
    \tau \to i\,\tilde{\sigma} \,,\qquad
    \sigma \to i\,\tilde{\tau} \,.
\end{equation}
While this transformation is immaterial for a relativistic theory, it has drastic effects for a non-relativistic theory such as the one on the lightcone gauge-fixed worldsheet~\cite{Arutyunov:2007tc}. It produces a new ``mirror'' model. In the cases considered thus far, the mirror model is related by an analytic continuation to the original one, and in this sense its integrability follows from that of the original theory.

In the case of strings in $AdS_3\times S^3\times T^4$ classical integrability holds when the background is supported by a mixture of Ramond-Ramond and Neveu-Schwarz-Neveu-Schwarz fluxes~\cite{Cagnazzo:2012se}.
The tree-level S~matrix of this model is integrable~\cite{Hoare:2013pma},
see also~\cite{Sundin:2014ema,Sundin:2016gqe},%
\footnote{%
In~\cite{Sundin:2014ema,Sundin:2016gqe} some one- and two-loops S~matrix elements have also been computed. However, as discussed in~\cite{Frolov:2021fmj}, some subtleties related to the presence of IR divergences may affect the results.
}
and in fact integrability seems to extend to all loops~\cite{Lloyd:2014bsa}, see also~\cite{Sfondrini:2014via} for a review of $AdS_3$ integrability.
Still, the worldsheet dynamics is quite peculiar for the mixed-flux theory, as it can already be seen in the near-BMN limit. As we will discuss in detail below, after gauge fixing we are left with an action for eight bosons and as many fermions.%
\footnote{%
To describe the full superstring action, we would need to start from the Green-Schwarz action and fix lightcone $\kappa$-gauge too, see \textit{e.g.}~\cite{Lloyd:2014bsa}.
}
At leading order in the near-BMN expansion we have a free theory, with a dispersion relation
\begin{equation}
\label{eq:dispersion}
    \omega(p,m) = \sqrt{p^2+ 2\, q\, m \, p +m^2} = \sqrt{(p+q\,m)^2+ (1-q^2)\,m^2}\,,
\end{equation}
where $p$ is the worldsheet momentum of a particle, the integer~$m$ is a certain combination of~$\mathfrak{u}(1)$ charges which characterizes the type of excitation,%
\footnote{The fundamental fields of the model consist of two bosons with $m=+1$, two with $m=-1$, and four with~$m=0$; additionally, particles with other~$m\in\mathbb{Z}$ can appear as bound states~\cite{Frolov:2023lwd}.} 
and $0\leq q\leq 1$ is a parameter that interpolates between the case of a pure-RR background ($q=0$) and pure-NSNS background ($q=1$).
In fact, in the bosonic action $q$ is the coefficient of the Kalb-Ramond field, and as we may have expected its presence breaks parity-invariance on the worldsheet.
It is worth noting that in the case of $q=0$ we have a kinematics similar to that of $AdS_5\times S^5$, while at $q=1$ all particles move at the speed of light.%
\footnote{%
This corresponds to the fact that at $q=1$ the action becomes that of a WZW model. This can be solved by CFT techniques~\cite{Maldacena:2000hw} as well as by integrability in lightcone gauge~\cite{Baggio:2018gct,Dei:2018mfl}, where many simplifications occur due to this chiral kinematics.}
The mirror transformation~\eqref{eq:doublewick_tau_sig}, when formulated in terms of momentum~$p$ and energy~$\omega$, gives
\begin{equation}
\label{eq:doublewick_p_om}
    p \to i\,\tilde{\omega}\,,\qquad
    \omega \to i\,\tilde{p}\,.
\end{equation}
As a result, in the mirror model the dispersion relation reads
\begin{equation}
    \tilde{\omega}(\tilde{p},m)=\sqrt{\tilde{p}^2+(1-q^2)\,m^2}+i\,q\,m\,,
\end{equation}
which is complex for the particles with~$m\neq0$.
Note that this model is parity-symmetric, but it is not invariant under time reversal. This is expected, as the double Wick rotation~\eqref{eq:doublewick_tau_sig} has swapped time and space: the original string model was not parity-invariant, but it was invariant under time-reversal, see~\eqref{eq:dispersion}.%
\footnote{Parity and time-reversal are observed at the level of the Lagrangian by sending $\sigma \to - \sigma$ and $\tau \to - \tau$, respectively. In principle, such symmetries could be incompatible with integrability at the quantum level and violated by counterterms necessary to avoid loop-level production amplitudes. It is however quite unlikely that this happens and we expect the all-loop quantum S-matrices to satisfy these symmetries as well. In this respect, the time-reversal invariance of the string model and parity-invariance of the mirror model can be taken as axioms for bootstrap construction of the S-matrix.}
By the same token, in light of~\eqref{eq:doublewick_tau_sig}, the appearance of complex terms in the mirror dispersion~$\tilde{\omega}(\tilde{p},m)$ is unsurprising; yet it is troubling, given the importance that the mirror model has in the determination of the spectrum as well as in the computation of three- and higher-point functions~\cite{Basso:2015zoa,Eden:2021xhe}.
It is also worth stressing that the issue remains in the all-loop S~matrix, though the dispersion relation and S-matrix elements are more complicated and cannot be expressed in terms of elementary functions for the mirror theory. 

The aim of this paper is to study the S matrix of the mirror theory at the tree level. In particular, we will perturbatively compute the two-to-two tree-level S~matrix for the processes that involve bosons, and check that the three-to-three processes factorise --- more precisely that there is no particle production in the six-point amplitudes (along the way, we rederive the tree-level S~matrix for the string model known from~\cite{Hoare:2013pma,Sundin:2014ema,Sundin:2016gqe}). 
We will also check whether the result of the perturbative computation is compatible with what is expected from the all-loop S~matrix, and discuss whether the presence of complex energies can lead to inconsistencies in the computation of L\"uscher corrections.

This paper is structured as follows. We begin by reviewing the construction of the perturbative worldsheet S~matrix in lightcone gauge in section~\ref{sec:string_theory}. This closely follows the reviews~\cite{Arutyunov:2009ga,Sfondrini:2014via}. We then repeat the discussions for the mirror model in section~\ref{sec:mirror_theory}. After having found a tree-level S-matrix for the mirror model, we compare our results with the non-perturbative computation and discuss the reality of the finite-size corrections. Finally, in section~\ref{sec:production} we check the absence of particle production in six-point processes for the string and mirror theory.
In appendix~\ref{Appendix_on_interacting_Hamiltonians_and_Lagrangians} we collect some bulky expressions for the Lagrangian, and in appendix~\ref{Appendix_on_production_terms} we do the same for certain scattering amplitudes.

\section{Scattering on the worldsheet for strings in \texorpdfstring{$AdS_3 \times S^3 \times T^4$}{AdS3xS3xT4}}
\label{sec:string_theory}

In this section, we review the derivation of the tree-level S matrix describing the scattering of excitations on the worldsheet of strings propagating on the target manifold AdS$_3 \times$ S$^3 \times $ T$^4$ supported by mixed Ramond-Ramond and Neveu-Schwarz Neveu-Schwarz flux. We focus on the purely bosonic sector of the string and we perform the derivation perturbatively in large string tension, see also~\cite{Arutyunov:2009ga,Sfondrini:2014via}.

\subsection{Bosonic strings with mixed flux backgrounds}

Let us consider a closed bosonic string of circumference $\stl$; its classical action is defined by
\begin{equation}
\label{starting_string_action}
S=\int\limits_{-\infty}^{+\infty} \de\tau \int\limits_{0}^{\stl} \de\sigma\, \mathcal{L}\,,\qquad
\mathcal{L}=-\frac{T}{2} \left[ \gamma^{ab}\partial_aX^{\mu}\partial_bX^{\nu}G_{\mu\nu}+\epsilon^{\alpha\beta} B_{\mu \nu} \partial_{\alpha}X^{\mu}\partial_{\beta}X^{\nu} \right] \, . 
\end{equation}
$T$ is the string tension, $\gamma^{\alpha \beta}=\sqrt{-h} h^{\alpha \beta}$ is the Weyl invariant metric on the worldsheet, $G_{\mu \nu}$ is the metric on the target manifold $AdS_3 \times S^3 \times  T^4$, $B_{\mu \nu}$ is the Kalb-Ramond field and $\epsilon^{\alpha \beta}$ is the Levi-Civita antisimmetric tensor having $\epsilon^{01}=+1$.
The  metric and B-field are given by
\begin{equation}
\label{first_expression_for_the_metric}
\begin{split}
\de s^2=& - \Bigl( \frac{4+z_1^2 + z_2^2}{4-z_1^2 - z_2^2} \Bigr)^2 \de t^2+\Bigl( \frac{4}{4-z_1^2 - z_2^2} \Bigr)^2 (\de z_1^2+\de z_2^2)\\
&+\Bigl( \frac{4-y_1^2 - y_2^2}{4+y_1^2 + y_2^2} \Bigr)^2 \de \phi^2+\Bigl( \frac{4}{4+y_1^2 + y_2^2} \Bigr)^2 ( \de y_1^2+\de y_2^2)\\
&+\sum_{j=5}^8 \de x_j  \de x_j\,,
\end{split}
\end{equation}
and
\begin{equation}
\label{first_expression_for_the_B_field}
\begin{aligned}
B=&  \frac{32q}{(4-z_1^2-z_2^2)^2} \left[z_1 \de z_2 \wedge \de t + z_2 \de t \wedge \de z_1 \right]\\
&+ \frac{32 q}{(4+y_3^2+y_4^2)^2} \left[y_3 \de y_4 \wedge \de  \phi + y_4 \de  \phi \wedge \de y_3 \right]\,,
\end{aligned}
\end{equation}
respectively.  In the formulae above we labeled by $(t, z_1, z_2)$ the coordinates on $AdS_3$, by $(\phi, y_1, y_2)$ the coordinates on $S^3$ and by $(x_5, x_6, x_7, x_8)$ the coordinates on $T^4$.
The B-field vanishes on $T^4$. In fact, $H=\de B$ is proportional to the volume form on~$AdS_3\times S^3$, with proportionality coefficient~$q$.

Both the metric and the B-field components do not depend on $t$ and $\phi$; therefore translations in $t$ and $\phi$ are manifest isometries to which we associate the conserved charges 
corresponding to the target-space energy and angular momentum on~$S^3$, respectively. We denote them by
\begin{equation}
\mathbf{H^{\text{t.s.}}}=- \int\limits_{0}^{\stl} \de\sigma\, P_t \,,\qquad \mathbf{J}= \int\limits_{0}^{\stl} \de\sigma\, P_\phi \,,
\end{equation}
where we introduced the momenta
\begin{equation}
\label{definition_of_conjugate_momenta}
P_\mu= \frac{\delta \mathcal{L}}{\delta (\dot X^{\mu})} = -T \Bigl( \gamma^{\tau \beta} G_{\mu \nu} + \epsilon^{\tau \beta} B_{\mu \nu} \Bigr) \partial_{\beta} X^\nu \,.
\end{equation}
Here and below we use the short-hand notation
\begin{equation}
\dot X^{\mu} \equiv \partial_{\tau} X^{\mu}\,,\qquad  \acute X^{\mu} \equiv \partial_{\sigma} X^{\mu}\,.
\end{equation}

We introduce the light-cone coordinates
\begin{equation}
x_-=\phi - t\,,\qquad x_+=a \phi + (1-a) t \,,
\end{equation}
with $a\in\mathbb{R}$ a free real parameter. We also introduce complex coordinates for the bosons,
\begin{equation}
\begin{split}
&z= \frac{1}{\sqrt{2}}(z_1 - i z_2) \hspace{4mm},\hspace{4mm} y= \frac{1}{\sqrt{2}}(y_1 - i y_2),\\
&u= \frac{1}{\sqrt{2}}(x_5 - i x_6) \hspace{4mm},\hspace{4mm} v= \frac{1}{\sqrt{2}}(x_7 - i x_8) \,.
\end{split}
\end{equation}
Then the metric and B-field components become
\begin{equation}
\begin{aligned}
&B_{\bar{z} +}=( B_{z +})^*=\frac{-4iq  z}{(2-|z|^2)^2}  \, , \qquad  B_{\bar{z}-}=(B_{z -})^*=\frac{4aiq  z}{(2-|z|^2)^2} \, , \\
&B_{\bar{y}+} = (B_{y +})^*=\frac{-4iq  y}{(2-|y|^2)^2} \, , \qquad  B_{\bar{y}-}=(B_{y -})^*=\frac{-4(1-a)iq  y}{(2-|y|^2)^2} \, .
\end{aligned}
\end{equation}
and the metric is
\begin{equation}
\begin{aligned}
&G_{++}=- \Bigl( \frac{2+|z|^2}{2-|z|^2} \Bigr)^2 + \Bigl( \frac{2-|y|^2}{2+|y|^2} \Bigr)^2  \, ,\\
&G_{--}= - a^2 \Bigl( \frac{2+|z|^2}{2-|z|^2} \Bigr)^2 + (1-a)^2 \Bigl( \frac{2-|y|^2}{2+|y|^2} \Bigr)^2 \, , \\
&G_{+-}=a \Bigl( \frac{2+|z|^2}{2-|z|^2} \Bigr)^2 + (1-a) \Bigl( \frac{2-|y|^2}{2+|y|^2} \Bigr)^2 \, ,\\
&G_{z\bar{z}}=\frac{4}{(2-|z|^2)^2} \, ,\qquad  G_{y\bar{y}}=\frac{4}{(2+|y|^2)^2} \, ,\qquad G_{u \bar{u}}=G_{v\bar{v}}=1 .
\end{aligned}
\end{equation}

To define the action~\eqref{starting_string_action} in terms of this new set of coordinates we invert~\eqref{definition_of_conjugate_momenta} first; in this manner we find $\dot X^{\mu}$ as a function of $P_\mu$ and $\acute X^{\mu}$:
\begin{equation}
T \gamma^{\tau \tau} \dot X^{\rho} = -T \gamma^{\tau \sigma} \acute X^{\rho} - T \epsilon^{\tau \sigma} G^{\rho \mu} B_{\mu \nu} \acute X^{\nu}-G^{\rho \mu} P_{\mu} \, .
\end{equation}
Then we plug this relation in~\eqref{starting_string_action} and we obtain
\begin{equation}
\label{starting_string_action_formula_2}
S=\int\limits_{-\infty}^{+\infty} \de\tau \int\limits_{0}^{\stl} \de\sigma \Bigl( P_{\mu} \dot{X}^\mu +\frac{\gamma^{\tau \sigma}}{\gamma^{\tau \tau}} C_1 + \frac{1}{2 T \gamma^{\tau \tau}} C_2\Bigr)
\end{equation}
where
\begin{subequations}
    \label{C1_C2_first_definition}
\begin{align}
\label{C1_first_definition}
 &C_1 = P_{\mu}\acute{X}^{\mu}\, ,\\
\label{C2_first_definition}
&C_2 = G^{\mu\nu}P_{\mu}P_{\nu} + T^2 G_{\mu\nu}\acute{X}^{\mu}\acute{X}^{\nu}+ 2 T G^{\mu\nu}B_{\nu\kappa} P_\mu \acute{X}^\kappa + T^2 G^{\mu\nu}B_{\mu\kappa} B_{\nu\lambda} \acute{X}^{\kappa}\acute{X}^{\lambda} \, .
\end{align}
\end{subequations}

To remove redundant degrees of freedom we set the uniform lightcone gauge~\cite{Arutyunov:2007tc,Arutyunov:2009ga}
\begin{equation}
x^+= \tau\,,\qquad P_-=1\,.
\end{equation}
Solving the Virasoro constraints $C_1=C_2=0$ in this gauge returns
\begin{equation}
\label{string_action_first_hamiltonian_formulation}
S=\int\limits_{-\infty}^{+\infty} \de\tau \int\limits_{0}^{\stl} \de\sigma \Bigl( P_{n} \dot{X}^n - \mathcal{H} \Bigl)
\end{equation}
where we defined the Hamiltonian density as $\mathcal{H}=-P_+$ and $n$ spans the indices of the physical fields (i.e. $n$ runs over all the spacetime indices apart from $+$ and $-$). 

It is worth remarking that after light-cone gauge fixing both the worldsheet length and the worldsheet Hamiltonian are related to the string target space energy and angular momentum; this is clear by the fact that
\begin{equation}
P_+=P_\phi+P_t \hspace{4mm}, \hspace{4mm} P_-= (1-a) P_\phi - a P_t \,,
\end{equation}
from which we can write
\begin{equation}
\begin{split}
&\mathbf{H^{\text{ws}}}=-\int\limits_{0}^{\stl} \de \sigma\,P_+=\mathbf{H^{\text{t.s.}}} - \mathbf{J} \,, \\
&\stl= \int\limits_{0}^{\stl} \de \sigma\, P_- = (1-a)  \mathbf{J} + a  \mathbf{H^{\text{t.s.}}} \,.
\end{split}
\end{equation}
The string size and the worldsheet Hamiltonian are therefore determined in terms of the target space energy and angular momentum.
The remaining longitudinal degree of freedom is~$X^-$. This is not entirely fixed by the gauge fixing, but the constraint~$C_1=0$ implies that
\begin{equation}
\label{Xminus_as_function_of_physical_fields}
\acute{X}^{-}= -P_n\acute{X}^n \,.
\end{equation}
If we require~$X^-$ to satisfy periodic boundary conditions, we have
\begin{equation}
    0=\int\limits_0^\stl \de\sigma\left(-P_n\acute{X}^n\right)=-\mathbf{P}^{\text{ws}}\,,
\end{equation}
where we have noted that the integrand is proportional to the charge density of the worldsheet momentum (\textit{i.e.}, the charge related to $\sigma\to\sigma+\text{const.}$).

In what follows we will be interested in the \textit{decompactification limit}~\cite{Arutyunov:2009ga}, that is $\stl \to\infty$. In this way, the worldsheet becomes a plane, and we can define asymptotic scattering states.
Let us now turn to the worldsheet Hamiltonian which is given by solving the quadratic constraint~$C_2=0$,
\begin{equation}
\label{Second_order_equation_to_determine_Pplus}
 a_1 \mathcal{H}^2 -a_2 \mathcal{H} +a_3 =0 \,.
\end{equation}
For later convenience, it is worth rescaling $\sigma\to T\sigma$ in the action, which makes  $C_2$ independent from~$T$. Then the gauge-fixed action scales as  
\begin{equation}
    S_{\text{g.f.}}=T \int\limits_{-\infty}^{+\infty}\de\tau \int\limits_{0}^{+\infty}\de\sigma\, \mathcal{L}_{\text{g.f.}}\,,
\end{equation}
where $\mathcal{L}_{\text{g.f.}}$ is independent from the tension. Then, a large-$T$ expansion is tantamount to an expansion in the number of fields. The explicit form of the coefficients of~\eqref{Second_order_equation_to_determine_Pplus} is then
\begin{equation}
\begin{split}
a_1=&\; G^{++} \,,\\
a_2=&\; 2 G^{+ -}+2 G^{+ +} B_{+ a} \acute{X}^{a} +2 G^{+ -} B_{- a} \acute{X}^{a} \,,\\
a_3=&\; G^{--}+ G^{ab} P_a P_b + G_{ab} \acute{X}^{a} \acute{X}^{b} + G_{--} P_a P_b \acute{X}^a \acute{X}^b+ 2 G^{- -} B_{- a} \acute{X}^{a}\\
&+2 G^{- +} B_{+ a} \acute{X}^{a}+ 2 G^{a c} B_{- c} P_a P_b \acute{X}^b+ G^{c d} B_{- c} B_{- d} P_a P_b \acute{X}^a \acute{X}^b\\
&+ G^{+ +} B_{+ a} B_{+ b} \acute{X}^{a} \acute{X}^{b}+  G^{- -} B_{- a} B_{- b} \acute{X}^{a} \acute{X}^{b}+2  G^{+ -} B_{+ a} B_{- b} \acute{X}^{a} \acute{X}^{b} \,.
\end{split}
\end{equation}
There are two possible solutions for $\mathcal{H}$; the physical one corresponds to the positive sign of the square root and for this choice the Hamiltonian goes to zero in the limit of null fields.
Let us define the expansion of the Hamiltonian in fields as 
\begin{equation}
\mathcal{H}=\mathcal{H}^{(2)}+\mathcal{H}^{(4)} +\mathcal{H}^{(6)} \dots
\end{equation}
where $\mathcal{H}^{(2)}$, $\mathcal{H}^{(4)}$ and $\mathcal{H}^{(6)}$ correspond to second-, fourth- and sixth-order powers of the fields.
The quadratic Hamiltonian is given by
\begin{multline}
\label{second_order_Hamiltonian_string_model}
    \mathcal{H}^{(2)}=P_{z}P_{\bar z}+P_{y}P_{\bar y}+ P_{u}P_{\bar u}+P_{v}P_{\bar v}\\
    +|\acute{z}|^2 + |\acute{y}|^2+|\acute{u}|^2+ |\acute{v}|^2   + (|z|^2+|y|^2)- iq( \bar z \acute{z}-z \acute{\bar z}+ \bar y \acute{y}-y \acute{\bar y})\,,
\end{multline}
corresponding to four complex bosons, two of which are coupled to a constant magnetic background field.
Higher-order expressions can be found in appendix~\ref{appendix:Lagrangian_string_model}.

\subsection{The free bosonic string theory}

The quadratic Lagragian corresponding to~\eqref{second_order_Hamiltonian_string_model} is
\begin{multline}
\label{Lagrangian_of_free_string_model}
    \mathcal{L}^{(2)}=\dot{z} \dot{\bar{z}}+\dot{y} \dot{\bar{y}}+\dot{u} \dot{\bar{u}}+\dot{v} \dot{\bar{v}}\\
    -\acute{z} \acute{\bar{z}} -\acute{y} \acute{\bar{y}} -\acute{u} \acute{\bar{u}} -\acute{v} \acute{\bar{v}}   - (z \bar{z}+ y \bar{y})+ iq( \bar z \acute{z}-z \acute{\bar z}+ \bar y \acute{y}-y \acute{\bar y}).
\end{multline}
Out of the four complex bosons of the model, $z$ and $y$ are massive, while $u$ and $v$ are massless. The latter come from the coordinates on $T^4$.
Defining $\Box \equiv \partial_0^2 - \partial_1^2$, the equations of motion of the massive and massless fields are
\begin{subequations}
\label{free_equations_of_motion_string_model_massive}
    \begin{align}
\label{free_equations_of_motion_string_model_massive_z}
&(\Box +1) z= 2 i q \partial_1 z \, ,  &&(\Box +1) \bar{z}= -2 i q \partial_1 \bar{z}\, ,\\
\label{free_equations_of_motion_string_model_massive_y}
&(\Box +1) y= 2 i q \partial_1 y \, ,  &&(\Box +1) \bar{y}= -2 i q \partial_1 \bar{y} \,,
\end{align}
\end{subequations}
and
\begin{equation}
\label{free_equations_of_motion_string_model_massless}
\Box  u= 0 \, ,\qquad \Box  \bar{u}= 0\, ,\qquad \Box  v= 0 \, ,\qquad \Box  \bar{v}= 0\,,
\end{equation}
respectively.
Labelling by $p$ the spatial component of the momentum the dispersion relations of the massive particles are 
\begin{equation}
\label{dispersion_relations_zy_and_barzbary}
\begin{split}
\omega_+(p)\equiv \omega_z(p)= \omega_y(p)=\sqrt{p^2+2q p+1 } \,,\\
\omega_-(p)\equiv \omega_{\bar{z}}(p)= \omega_{\bar{y}}(p)=\sqrt{p^2-2q p+1 } \, ,
\end{split}
\end{equation}
while the dispersion relations for the massless particles are
\begin{equation}
\label{dispersion_relation_masslss_in_string_model}
\omega_{\circ}(p)\equiv \omega_u(p)= \omega_v(p)=\omega_{\bar{u}}(p)= \omega_{\bar{v}}(p)=|p| \,.
\end{equation}
As we did in the introduction, see~\eqref{eq:dispersion}, we can write
\begin{equation}
\label{eq:string_dispersion_relation_carot}
    \omega(p,m)= \sqrt{p^2+2 \, q \, m \, p+m^2}\,,
\end{equation}
where $m$ is the sum of the spins along $AdS_3$ and $S^3$ of the fields; this definition would nicely generalise to the fermions too~\cite{Lloyd:2014bsa}.
From equation \eqref{dispersion_relations_zy_and_barzbary} we note that the energies of the massive particles are real and positive for all values of $p$ only if $q \in \mathbb{R}$ and $q^2\leq 1$, which is indeed the regime where our model is well defined. In this region of the $q$-parameter space, the Lagrangian of the string model is real and it is possible to compute amplitudes by applying standard perturbation theory.

Let us briefly review how the fields $z$ and $\bar{z}$ can be quantized. The quantisation for the remaining fields can be performed similarly. 
Solutions for $z$ and $\bar{z}$ satisfying~\eqref{free_equations_of_motion_string_model_massive_z} are provided by
\begin{equation}
\label{z_and_barz_in_terms_of_creation_and_annihilation_operators}
\begin{aligned}
&z(\tau,\sigma) = \int \frac{dp}{\sqrt{(2\pi)}}\biggl[\frac{e^{-i(\omega_+(p)\tau - p\sigma) }}{\sqrt{2\omega_+(p)}}a_z(p) + \frac{e^{i(\omega_-(p)\tau - p\sigma) }}{\sqrt{2\omega_-(p)}}a^\dagger
_{\Bar{z}}(p)\biggr]\, ,\\
&\Bar{z}(\tau,\sigma) = \int \frac{dp}{\sqrt{(2\pi)}}\biggl[\frac{e^{-i(\omega_-(p)\tau - p\sigma) }}{\sqrt{2\omega_-(p)}}a_{\Bar{z}}(p) + \frac{e^{i(\omega_+(p)\tau - p\sigma) }}{\sqrt{2\omega_+(p)}}a^\dagger
_z(p)\biggr]\, .
\end{aligned}
\end{equation}
In the equations above $a_z$ and $a_z^\dagger$ correspond to the annihilation and creation operators associated with particles of type $z$ while $a_{\bar{z}}$ and $a_{\bar{z}}^\dagger$ correspond to the annihilation and creation operators associated with particles of type $\bar{z}$.
The canonical quantization conditions require
\begin{equation}
\label{canonical_quantisation_conditions_for_z_string}
[z(\tau, \sigma), P_z(\tau, \sigma')] = i \delta(\sigma-\sigma') \hspace{3mm} , \hspace{3mm}[\bar{z}(\tau, \sigma), P_{\bar{z}}(\tau, \sigma')] = i \delta(\sigma-\sigma') \, 
\end{equation}
where at the leading order 
\begin{equation}
P_z = \frac{\partial \mathcal{L}^{(2)}}{\partial \dot{z}}=\dot{\bar{z}} \hspace{4mm} \text{and} \hspace{4mm} P_{\bar{z}} = \frac{\partial \mathcal{L}^{(2)}}{\partial \dot{\bar{z}}}=\dot{z} \,.
\end{equation}
From the canonical quantization conditions in~\eqref{canonical_quantisation_conditions_for_z_string} we derive the commutation relations between creation and annihilation operators:
\begin{equation}
\label{commutation_relations_between_a_and_dagger_a}
[a_z(p), a_z^\dagger (k)] = \delta(p-k) \hspace{3mm}, \hspace{3mm}[a_{\bar{z}}(p), a_{\bar{z}}^\dagger (k)] = \delta(p-k) \, .
\end{equation}
As usual, we define the vacuum to be the state $| 0\rangle$ annihilated by all the annihilation operators.
The ket and bra associated with double-particle states of the $z$ field can then be written as
\begin{equation}
\label{ket_double_particle_state_z_field}
| z(k) z(p) \rangle \equiv a_z^\dagger (k) a_z^\dagger (p)| 0\rangle
\end{equation}
and
\begin{equation}
\label{bra_double_particle_state_z_field}
\langle z(k') z(p') |  \equiv \langle 0 | a_z (k') a_z (p') .
\end{equation}
Using the fact that $a_z(p) | 0\rangle =0$ and $\langle 0 |a^\dagger_z(p)=0$ $\forall \ p$ together with the commutation relations in~\eqref{commutation_relations_between_a_and_dagger_a} then we obtain
\begin{equation}
\label{scalar_product_between_states}
\langle z(k') z(p') | z(k) z(p) \rangle = \delta( k' - k) \delta( p' - p) + \delta( k' - p) \delta( p' - k) \,.
\end{equation}
Similar relations can be obtained for all states of the theory. From the scalar product~\eqref{scalar_product_between_states} we see that the states, as defined in~\eqref{ket_double_particle_state_z_field} and~\eqref{bra_double_particle_state_z_field}, have the correct normalisation and can be directly plugged into the scattering matrix of the interacting theory to compute amplitudes.

\subsection{Tree-level scattering in the string model}

The S-matrix of the theory is defined by
\begin{equation}
\mathbb{S}= \mathbb{I} + i\,  \mathbb{T} \, ,
\end{equation}
where the operator $\mathbb{T}$ takes into account the interactions and can schematically be written as
\begin{equation}
\mathbb{T}= -\int \de\tau \de\sigma \, :\mathcal{H}^{(4)}: -\int \de\tau \de\sigma \, :\mathcal{H}^{(6)}: +\frac{i}{2} \Bigl( \int \de\tau \de\sigma \, :\mathcal{H}^{(4)}: \Bigl)^2+\dots \, .
\end{equation}
The ellipses in the expression above contain interaction terms
contributing to tree-level scattering processes with more than six external legs, which will not be considered in this paper. 
We will see that the S-matrices of the string and mirror models are purely elastic, with no production at the tree level~\cite{Hoare:2013pma,Lloyd:2014bsa}. For this reason, we will often label their elements by
\begin{equation}
\label{definition_scalars_S}
\mathbb{S}\ket{A(p) B(p')} = \sr_{AB}(p, p') \ket{B(p') A(p)} \,,
\end{equation}
where $A$ and $B$ are generic particles carrying momenta $p$ and $p'$. Similarly, for the interaction part we can write
\begin{equation}
\label{definition_scalars_T}
\mathbb{T}\ket{A(p) B(p')} = \tr_{AB}(p, p') \ket{B(p') A(p)} \,.
\end{equation}

\begin{figure}[t]
\begin{center}
    \begin{tikzpicture}
\tikzmath{\y=2.8;}

\filldraw[black] (0.43*\y,-1*\y)  node[anchor=west] {\scriptsize{$\bullet$}};
\draw[] (0.2*\y,-1.3*\y) -- (0.8*\y,-0.7*\y);
\draw[] (0.8*\y,-1.3*\y) -- (0.2*\y,-0.7*\y);
\draw[->] (0.1*\y,-1.3*\y) -- (0.3*\y,-1.1*\y);
\draw[->] (0.8*\y,-0.6*\y) -- (0.6*\y,-0.8*\y);
\draw[->] (0.2*\y,-0.6*\y) -- (0.4*\y,-0.8*\y);
\draw[->] (0.9*\y,-1.3*\y) -- (0.7*\y,-1.1*\y);
\filldraw[black] (-0.1*\y,-1.2*\y)  node[anchor=west] {\scriptsize{$X(p)$}};
\filldraw[black] (0.82*\y,-1.2*\y)  node[anchor=west] {\scriptsize{$X(p')$}};
\filldraw[black] (0.7*\y,-0.55*\y)  node[anchor=west] {\scriptsize{$\bar{X}(k)$}};
\filldraw[black] (0.1*\y,-0.55*\y)  node[anchor=west] {\scriptsize{$\bar{X}(k')$}};

\filldraw[black] (1.1*\y,-1.4*\y)  node[anchor=west] {\small{$\begin{aligned}
=& \mathscr{S}_X i \Bigl(2   p'_1  k'_1  +2 p'_1 k_1+2 p_1 k'_1  +2 p_1  k_1\\
&- q \bigl[ (k_1+k_1') (p_0 p_0'-p_1 p'_1 -1)-(p_1+p_1') (k_0 k_0'-k_1 k'_1 -1) \bigl] \Bigl)\\
&+i (2a -1) \Bigl( 2-2 (p_0 p_0'-p_1 p_1') (k_0 k_0'-k_1 k_1')\\
&-q \bigl[ (k_1+k_1') (p_0 p_0'-p_1 p'_1 +1)-(p_1+p_1') (k_0 k_0'-k_1 k'_1 +1) \bigl] \Bigl)
\end{aligned}$}};


\filldraw[black] (0.43*\y,-1*\y-1.6*\y)  node[anchor=west] {\scriptsize{$\bullet$}};
\draw[] (0.2*\y,-1.3*\y-1.6*\y) -- (0.8*\y,-0.7*\y-1.6*\y);
\draw[] (0.8*\y,-1.3*\y-1.6*\y) -- (0.2*\y,-0.7*\y-1.6*\y);
\draw[->] (0.1*\y,-1.3*\y-1.6*\y) -- (0.3*\y,-1.1*\y-1.6*\y);
\draw[->] (0.8*\y,-0.6*\y-1.6*\y) -- (0.6*\y,-0.8*\y-1.6*\y);
\draw[->] (0.2*\y,-0.6*\y-1.6*\y) -- (0.4*\y,-0.8*\y-1.6*\y);
\draw[->] (0.9*\y,-1.3*\y-1.6*\y) -- (0.7*\y,-1.1*\y-1.6*\y);
\filldraw[black] (-0.1*\y,-1.1*\y-1.6*\y)  node[anchor=west] {\scriptsize{$y(p)$}};
\filldraw[black] (0.75*\y,-1.1*\y-1.6*\y)  node[anchor=west] {\scriptsize{$z(p')$}};
\filldraw[black] (0.7*\y,-0.55*\y-1.6*\y)  node[anchor=west] {\scriptsize{$\bar{y}(k)$}};
\filldraw[black] (0.1*\y,-0.55*\y-1.6*\y)  node[anchor=west] {\scriptsize{$\bar{z}(k')$}};

\filldraw[black] (1*\y,-1*\y-2.2*\y)  node[anchor=west] {\small{$\begin{aligned}
&=i \Bigl( p_0k_0+p_1k_1-p_0'k_0'-p_1'k_1'\\
&-\frac{q}{2}\bigl[(p_1'k_0'+k_1'p_0')( k_0-p_0)+(p_1k_0 + k_1p_0)(p_0'-k_0')\\
&+(1+p_0'k_0'+p_1'k_1')(p_1-k_1)-(1+p_0k_0 +p_1k_1)(p_1'-k_1')\bigl] \Bigl)\\
&+ i (2a-1) \Bigl( 1 + (p'_1 k'_0 +k'_1 p'_0) (p_1 k_0 +k_1 p_0) - (p'_0 k'_0 +p'_1 k'_1) (p_0 k_0 +p_1 k_1)\\
&-\frac{q}{2} \bigl[ (k_1'-p_1') (p_0 k_0+p_1 k_1+1) + (k_1-p_1) (p'_0 k'_0+p'_1 k'_1+1)\\
&+(p_0-k_0) (p'_0 k'_1+p'_1 k'_0) +(p'_0-k'_0) (p_0 k_1+p_1 k_0) \bigl] \Bigl)
\end{aligned}$}};


\draw[] (0.2*\y,-1.3*\y-3.6*\y) -- (0.8*\y,-0.7*\y-3.6*\y);
\draw[] (0.8*\y,-1.3*\y-3.6*\y) -- (0.2*\y,-0.7*\y-3.6*\y);
\draw[->] (0.1*\y,-1.3*\y-3.6*\y) -- (0.3*\y,-1.1*\y-3.6*\y);
\draw[->] (0.8*\y,-0.6*\y-3.6*\y) -- (0.6*\y,-0.8*\y-3.6*\y);
\draw[->] (0.2*\y,-0.6*\y-3.6*\y) -- (0.4*\y,-0.8*\y-3.6*\y);
\draw[->] (0.9*\y,-1.3*\y-3.6*\y) -- (0.7*\y,-1.1*\y-3.6*\y);
\filldraw[black] (-0.1*\y,-3.6*\y-1.2*\y)  node[anchor=west] {\scriptsize{$\mu(p)$}};
\filldraw[black] (0.82*\y,-1.2*\y-3.6*\y)  node[anchor=west] {\scriptsize{$X(p')$}};
\filldraw[black] (0.7*\y,-0.55*\y-3.6*\y)  node[anchor=west] {\scriptsize{$\bar{\mu}(k)$}};
\filldraw[black] (0.1*\y,-0.55*\y-3.6*\y)  node[anchor=west] {\scriptsize{$\bar{X}(k')$}};

\filldraw[black] (0.43*\y,-1*\y-3.6*\y)  node[anchor=west] {\scriptsize{$\bullet$}};

\filldraw[black] (1.1*\y,-1*\y-4*\y)  node[anchor=west] {\small{$\begin{aligned}
=& \mathscr{S}_X i \Bigl( (p_0k_0+p_1k_1)\\
&-\frac{q}{2} \bigl[(p_1k_0 + k_1p_0)(p_0'-k_0')-(p_0k_0+p_1 k_1)(p_1'-k'_1) \bigl] \Bigl)\\
&+i (2a-1) \Bigl( (p_1' k_0' + p_0' k_1')(p_1 k_0 + p_0 k_1)-(p_0' k_0' + p_1' k_1')(p_0 k_0 + p_1 k_1)\\
&-\frac{q}{2} \bigl[ (p_0 k_1 + p_1 k_0) (p_0'-k_0') - (p_0 k_0 + p_1 k_1) (p_1'-k_1') \bigl]\Bigl)
\end{aligned}$}};
\end{tikzpicture}
\begin{tikzpicture}
\tikzmath{\y=3;}

\end{tikzpicture}
\end{center}
\caption{Massive and massive-massless interaction vertices of the bosonic string model with mixed three-form flux. The field $X$ can either be $z$ and the field $\mu$ can either be $u$ or $v$.. We label by $p_0$ and $p_1$ the off-shell temporal and spatial components of the momentum $p$. The same definition applies to the momenta $p'$, $k$ and $k'$. The function $\mathscr{S}_X$ is defined in~\eqref{eq:sign_X_z_y}.
}
\label{image_vertex_interactions_string_theory_massive}
\end{figure}

\begin{figure}[t]
\begin{center}
    \begin{tikzpicture}
\tikzmath{\y=2.8;}

\filldraw[black] (0.43*\y,-1*\y)  node[anchor=west] {\scriptsize{$\bullet$}};
\draw[] (0.2*\y,-1.3*\y) -- (0.8*\y,-0.7*\y);
\draw[] (0.8*\y,-1.3*\y) -- (0.2*\y,-0.7*\y);
\draw[->] (0.1*\y,-1.3*\y) -- (0.3*\y,-1.1*\y);
\draw[->] (0.8*\y,-0.6*\y) -- (0.6*\y,-0.8*\y);
\draw[->] (0.2*\y,-0.6*\y) -- (0.4*\y,-0.8*\y);
\draw[->] (0.9*\y,-1.3*\y) -- (0.7*\y,-1.1*\y);
\filldraw[black] (-0.1*\y,-1.2*\y)  node[anchor=west] {\scriptsize{$\mu(p)$}};
\filldraw[black] (0.82*\y,-1.2*\y)  node[anchor=west] {\scriptsize{$\mu(p')$}};
\filldraw[black] (0.7*\y,-0.55*\y)  node[anchor=west] {\scriptsize{$\bar{\mu}(k)$}};
\filldraw[black] (0.1*\y,-0.55*\y)  node[anchor=west] {\scriptsize{$\bar{\mu}(k')$}};

\filldraw[black] (1.1*\y,-1*\y)  node[anchor=west] {\small{$\begin{aligned}
=&-2 i (2 a - 1) (p_0 p'_0-p_1 p'_1) (k_0 k'_0-k_1 k'_1)
\end{aligned}$}};


\filldraw[black] (0.43*\y,-1*\y-1.2*\y)  node[anchor=west] {\scriptsize{$\bullet$}};
\draw[] (0.2*\y,-1.3*\y-1.2*\y) -- (0.8*\y,-0.7*\y-1.2*\y);
\draw[] (0.8*\y,-1.3*\y-1.2*\y) -- (0.2*\y,-0.7*\y-1.2*\y);
\draw[->] (0.1*\y,-1.3*\y-1.2*\y) -- (0.3*\y,-1.1*\y-1.2*\y);
\draw[->] (0.8*\y,-0.6*\y-1.2*\y) -- (0.6*\y,-0.8*\y-1.2*\y);
\draw[->] (0.2*\y,-0.6*\y-1.2*\y) -- (0.4*\y,-0.8*\y-1.2*\y);
\draw[->] (0.9*\y,-1.3*\y-1.2*\y) -- (0.7*\y,-1.1*\y-1.2*\y);
\filldraw[black] (-0.1*\y,-1.2*\y-1.2*\y)  node[anchor=west] {\scriptsize{$u(p)$}};
\filldraw[black] (0.82*\y,-1.2*\y-1.2*\y)  node[anchor=west] {\scriptsize{$v(p')$}};
\filldraw[black] (0.7*\y,-0.55*\y-1.2*\y)  node[anchor=west] {\scriptsize{$\bar{u}(k)$}};
\filldraw[black] (0.1*\y,-0.55*\y-1.2*\y)  node[anchor=west] {\scriptsize{$\bar{v}(k')$}};

\filldraw[black] (1*\y,-1*\y-1.2*\y)  node[anchor=west] {\small{$\begin{aligned}
&=(2a - 1) i\bigl[(p_1'k_0' + p_0'k_1')(p_1k_0 + p_0k_1) -(p_0'k_0' + p_1'k_1')(p_0k_0 + p_1k_1)\bigl]
\end{aligned}$}};
\end{tikzpicture}
\end{center}
\caption{Massless interaction vertices of the bosonic string model with mixed three-form flux. We follow the same conventions used in figure~\ref{image_vertex_interactions_string_theory_massive}.
}
\label{image_vertex_interactions_string_theory_massless}
\end{figure}

An easy way to compute the tree-level scattering amplitudes is to read them off from the Lagrangian by extracting the interacting vertices using standard Feynman rules. The fourth-order interacting Lagrangian of the string model is summarised in appendix~\ref{Appendix_on_interacting_Hamiltonians_and_Lagrangians}, while the associated vertices are shown in figures~\ref{image_vertex_interactions_string_theory_massive} and~\ref{image_vertex_interactions_string_theory_massless}, where we defined
\begin{equation}
\label{eq:sign_X_z_y}
\mathscr{S}_X = 
\begin{cases}
&+1 \quad \text{if} \quad  X=z \,,\\
&-1 \quad \text{if}  \quad X=y \,.
\end{cases}
\end{equation}
For convention, we assume all the momenta to be incoming into the vertices. 
Tree-level two-to-two amplitudes are proportional to these vertices. In addition, to obtain the amplitudes, we need to multiply by the following factors:
\begin{equation}
\label{extra_multiplicative_factors_to_be_added_in_amplitudes}
\begin{split}
&\text{for each external particle $i$:} \hspace{10mm} \frac{1}{\sqrt{2 \omega_i}} \,,\\
&\text{an overall multiplicative factor:} \hspace{6mm} \delta\Big(\sum_{i} p^{\text{in}}_i -\sum_{i} p^{\text{out}}_i\Big)\; \delta\Big(\sum_{i} \omega^{\text{in}}_i -\sum_{i} \omega^{\text{out}}_i\Big) .
\end{split}\end{equation}
The dispersion relations $\omega_i$ for the energies as functions of the momenta are written in eqs.~\eqref{dispersion_relations_zy_and_barzbary} and~\eqref{dispersion_relation_masslss_in_string_model}.

Even if in figures~\ref{image_vertex_interactions_string_theory_massive} and~\ref{image_vertex_interactions_string_theory_massless} the vertices are written with the convention that all the particles are incoming, by crossing symmetry it is always possible to transform an incoming particle into an outgoing antiparticle with opposite energy and momentum and computing in this manner all the two-to-two tree-level scattering amplitudes. From these vertices, it is also possible to check that the processes in which incoming and outgoing particles are of different types cancel and the tree-level S-matrix is therefore purely elastic (this is only true when dropping the boson-bosons--fermion-fermion processes~\cite{Lloyd:2014bsa}). 

Let us consider for example the vertex in the second row of figure~\ref{image_vertex_interactions_string_theory_massive}.
A tree-level amplitude associated with this vertex is
\begin{equation}
\label{eq:ybary_to_zbarz}
y(p) + \bar{y}(k) \to z(k') + \bar{z}(p') \, .
\end{equation}
Let us check that this amplitude vanishes at the tree level.
To avoid long expressions we perform the check in the gauge $a=\frac{1}{2}$, which yields particularly simple expressions; by crossing symmetry we can write
\begin{equation}
\label{ybary_into_zbarz_tree_level_amplitude}
\begin{split}
&\langle \bar{z}(p') z(k') | i \mathbb{T} | y(p) \bar{y}(k) \rangle\\
&\qquad=\langle 0 | i \mathbb{T} | y(p) \bar{y}(k) z(-p') \bar{z}(-k') \rangle\\
&\qquad=i\,\mathcal{N}(p,p')\, \biggl(  (p_0k_0+p_1k_1-p_0'k_0'-p_1'k_1')\\
&\qquad\qquad-\frac{iq}{2}\Big[{(p_1'k_0'+k_1'p_0')( k_0-p_0)+(p_1k_0 + k_1p_0)(-p_0'+k_0')}\\
&\qquad\qquad\qquad+(1+p_0'k_0'+p_1'k_1')(p_1-k_1)-(1+p_0k_0 +p_1k_1)(-p_1'+k_1')\Big] \biggl) \, ,
\end{split}
\end{equation}
where $\mathcal{N}(p,p')$ is the overall factor coming from multiplying by the terms in~\eqref{extra_multiplicative_factors_to_be_added_in_amplitudes} which we leave implicit. There are two branches of kinematics conserving the overall energy and momentum: 
\begin{equation}
\begin{split}
&1. \ \{k_1'=p_1, \, p_1'=k_1\} \, , \ \hspace{16.5mm} \text{with associated energies} \  \{k_0'=p_0, \, p_0'=k_0\} \, ;\\
&2. \ \{k_1'=k_1-2q, \, p_1'=p_1 + 2q\} \, , \ \text{with associated energies} \  \{k_0'=k_0, \, p_0'=p_0\} \, .
\end{split}
\end{equation}
Substituting any of the solutions above in~\eqref{ybary_into_zbarz_tree_level_amplitude} and taking the external particles on-shell we obtain a null result.
For $a\ne \frac{1}{2}$ we obtain an additional term in the amplitude, linear in $a - \frac{1}{2}$. It is easy to verify that this additional term cancels separately and this inelastic tree-level amplitude vanishes in all gauges.
It can be similarly proven that all the other two-to-two inelastic processes are zero and the tree-level S-matrix is therefore reflectionless.
We also remark that creation and annihilation processes of type $1 \to 3$ and $3 \to 1$ respectively are always forbidden for kinematical reasons. Note indeed that even though the process
\begin{equation}
\label{eq:creationoneto3}
y \to y+z+\bar{z}
\end{equation}
is connected to~\eqref{eq:ybary_to_zbarz} by crossing there are no physical solutions to the overall energy-momentum conservation for~\eqref{eq:creationoneto3}.

More interesting is the evaluation of tree-level amplitudes associated with elastic processes. To compute correctly these amplitudes we recall that we study the evolution of a state subject to no interaction in the far past ($\tau \to -\infty$) into a state subject to no interaction in the far future ($\tau \to +\infty$).
The requirement that each state is free for $\tau \to -\infty$ implies that particles composing incoming states need to be located from the left to the right on the
spatial line following the decreasing order of their velocities; in the two-body scattering, this is satisfied if we choose $v>0>v'$, where $v$ and $v'$ are the velocities of the particles incoming from the left and from the right respectively. Since the velocity of a string particle is defined as
\begin{equation}
\label{eq:velocity_carot}
v= \frac{\partial \omega}{\partial p} \,,
\end{equation}
with $\omega$ given by~\eqref{eq:string_dispersion_relation_carot}, then the condition $v>0>v'$ implies $p> -m q$ for a particle incoming from the left and $p'< -m' q$ for a particle incoming from the right, where $m$ and $m'$ are the quantum numbers of the incoming particles and can be either $0$ or $\pm 1$. This is important for the scattering to be well-defined perturbatively and turns out to be an important condition to remove absolute values arising from expressing the Dirac delta function of the overall energy-momentum conservation in terms of the spatial momenta.
With this convention, all the tree-level S-matrix elements of the string model can be computed. 
It is worth singling out the antisymmetric combination~\cite{
Arutyunov:2005hd} 
\begin{equation}
\label{eq:CDD_TbarT}
\Phi_{ij}(p,p') \equiv \frac{2a-1}{2}\,  \big[\omega_i(p)\,p'- \omega_j(p')\,p\big]\,, 
\end{equation}
where $\omega_i$ is the dispersion $\omega_+$, $\omega_-$ or $\omega_\circ$ depending on the type of particle under consideration. This is precisely the CDD factor that one would get from a $T\bar{T}$ deformation, which appears naturally in the uniform lightcone gauge~\cite{Baggio:2018gct}. 
Then the S-matrix elements of the string model are:
\paragraph{Massive-massive}
\begin{equation}
\label{eq:massive_massive_string_S_matrices}
\begin{aligned}
& \mathbb{T}\ket{X_{\pm}(p)X_{\pm}(p')} = \Bigl( -\mathscr{S}_X  \frac{(p+p')(\omega_{\pm} \, p'+\omega_{\pm}' p)}{2(p -p')}  +\Phi_{\pm\pm}(p, p') \Bigl)\ket{X_{\pm}(p')X_{\pm}(p)} ,\\
& \mathbb{T}\ket{X_{\pm}(p)X_{\mp}(p')} = \Bigl( -\mathscr{S}_X  \frac{(p-p')(\omega_{\pm} \, p'+\omega_{\mp}' p)}{2(p +p')}  +\Phi_{\pm\mp}(p, p') \Bigl)\ket{X_{\mp}(p')X_{\pm}(p)} ,\\
& \mathbb{T}\ket{z_{\pm}(p)y_{\pm}(p')} = \Bigl( \frac{1}{2}(\omega_{\pm} \, p' + \omega_{\pm}' \, p)  + \Phi_{\pm\pm}(p, p') \Bigl) \ket{y_{\pm}(p') z_{\pm}(p)} ,\\
& \mathbb{T}\ket{z_{\pm}(p)y_{\mp}(p')} = \Bigl( \frac{1}{2}(\omega_{\pm} \, p' + \omega_{\mp}' \, p)  + \Phi_{\pm\mp}(p, p') \Bigl) \ket{y_{\mp}(p') z_{\pm}(p)} ,
\end{aligned}
\end{equation}
\paragraph{Mixed-mass}
\begin{equation}
\label{eq:massive_massless_string_S_matrices}
\begin{aligned}
&\mathbb{T}\ket{X_{\pm}(p) U (p')} = \Bigl( \mathscr{S}_X \frac{1}{2}(\omega_{\pm} \, p'+\omega_{\circ}' \, p)+ \Phi_{\pm\circ}(p, p') \Bigr)\ket{ U(p') X_{\pm}(p)} ,\\
&\mathbb{T}\ket{U(p) X_{\pm} (p')} = \Bigl( -\mathscr{S}_X \frac{1}{2}(\omega_{\circ} \, p'+\omega_{\pm}' \, p)+ \Phi_{\circ\pm}(p, p') \Bigl) \ket{ X_{\pm}(p') U(p)} .
\end{aligned}
\end{equation}
\paragraph{Massless}
\begin{equation}
\label{eq:massless_massless_string_S_matrices}
\mathbb{T} \ket{U(p) V(p')} = \Phi_{\circ\circ}(p, p') \ket{V(p') U(p)} \,.
\end{equation}
In the formulas above, $X_+$ can either be $z$ or $y$, $X_-$ can either be $\bar{z}$ or $\bar{y}$ and $U$ and $V$ can either be $u$, $v$, $\bar{u}$ or $\bar{v}$. The sign function $\mathscr{S}_X$ is defined in~\eqref{eq:sign_X_z_y}. 
Moreover we adopted the shortcut notation $\omega_{\pm}(p) = \omega_{\pm}$, $\omega_{\pm}(p') = \omega_{\pm}'$, $\dots$, where the dispersion relations are written in~\eqref{dispersion_relations_zy_and_barzbary} and~\eqref{dispersion_relation_masslss_in_string_model}. In the gauge $a=\frac{1}{2}$ it holds that $\Phi_{ij}=0$ and the S-matrix elements simplify. The same conventions will be used for the mirror model. Our S-matrix elements for the string model are equivalent to the ones early found in~\cite{Hoare:2013pma,Sundin:2014ema,Sundin:2016gqe}.

Even though these tree-level S-matrices have been computed in the region of physical incoming particles having $p> -m q$ and $p'< -m' q$ they can be analytically continued to all values of $p$ and $p'$ and can be thought as functions of two complex parameters.
The remaining elements not provided in the expressions above can be computed by braiding unitarity.

It is important to highlight that the different S-matrix elements of the string model are connected by certain symmetries. 
Important symmetries are \textit{braiding unitarity} and \textit{crossing}, which at the tree level take the form
\begin{equation}
\label{eq:braiding_unit_crossing_string}
\begin{split}
\text{braiding unitarity:} \ \ \ &\tr_{AB}(p, q) = -\tr_{BA}(q, p) \,;\\
\text{crossing:} \ \ \ &\tr_{AB}(p, q)=-\tr_{A \bar{B}}(p, -q)=-\tr_{\bar{A} B}(-p, q)=\tr_{\bar{A} \bar{B}}(-p, -q)\,.
\end{split}
\end{equation}
In~\eqref{eq:braiding_unit_crossing_string} $p$ and $q$ need to be thought of as two-dimensional vectors containing both the energy and the spatial momentum of the particles.
An additional symmetry of the \textit{string model} is:
\begin{equation}
\text{CP:} \ \ \ \tr_{AB}(\{p_0, p_1\}, \{q_0, q_1\})=\tr_{\bar{B} \bar{A}}(\{q_0, -q_1\}, \{p_0, -p_1\}) \,.
\end{equation}
This symmetry corresponds to a combination of parity and charge conjugation and is due to the invariance of the string Lagrangian under a simultaneous transformation $\sigma \to -\sigma$ and $q \to -q$.

\section{Mirror theory}
\label{sec:mirror_theory}

As extensively described in~\cite{Arutyunov:2007tc} a necessary step to derive the finite-volume spectrum of the theory is to study the mirror model, defined by a double Wick rotation of time and space\footnote{Strictly speaking the mirror transformation should map 
$\tau \to -i \tilde{\sigma}$ and $\sigma \to i \tilde{\tau}$; however since the mirror theory obtained in this manner is parity invariant (i.e. the mirror Lagrangian does not change if we send $\tilde{\sigma} \to -\tilde{\sigma}$) then we can additionally reverse the space direction.}:
\begin{equation}
\label{eq:mirror_transformation_on_coordinates}
\tau \to i \tilde{\sigma} \hspace{3mm}, \hspace{3mm} \sigma \to i \tilde{\tau} \, .
\end{equation}
 The theory analytically continued to the mirror region is therefore obtained by substituting
\begin{equation}
\label{eq:map_string_mirror_derivatives}
\partial_\tau \to -i \partial_{\tilde{\sigma}} \hspace{3mm} \text{and} \hspace{3mm} \partial_\sigma \to -i \partial_{\tilde{\tau}}
\end{equation}
in the gauge fixed Lagrangian of the string model and the energies and momenta of the mirror particles can be obtained from the energies and momenta of the string particles by replacing
\begin{equation}
\label{eq:string_mirror_energies_and_momenta}
\tilde{p}=-i \omega \quad, \quad \tilde{\omega}= -i p \,.
\end{equation}

In this section, we derive the Lagrangian and S-matrix of the mirror model and we discuss their properties.

\subsection{Mirror transformation and free mirror theory}
\label{subsec_mirror_theory}

By applying the map in~\eqref{eq:map_string_mirror_derivatives} to equation~\eqref{Lagrangian_of_free_string_model} we obtain the following free Lagrangian for the mirror theory
\begin{multline}
    \label{Lagrangian_of_free_mirror_model}
    \mathcal{L}^{(2)}_m=\dot{z} \dot{\bar{z}}+\dot{y} \dot{\bar{y}}+ \dot{u} \dot{\bar{u}}+\dot{v} \dot{\bar{v}}\\
    -\acute{z} \acute{\bar{z}} -\acute{y} \acute{\bar{y}} -\acute{u} \acute{\bar{u}} -\acute{v} \acute{\bar{v}}   - (z \bar{z}+ y \bar{y})+ q \bigl(\bar{z}\dot{z}-z\dot{\bar{z}} + \bar{y}\dot{y}-y\dot{\bar{y}} \bigl)
\end{multline}
While the EOMs for the massless fields are invariant under the mirror transformation, the EOMs for the massive field become
\begin{subequations}
\label{free_equations_of_motion_mirror_model_massive}
    \begin{align}
\label{free_equations_of_motion_mirror_model_massive_z}
&(\Box +1) z= 2 q \partial_0 z \, ,  &&(\Box +1) \bar{z}= -2 q \partial_0 \bar{z}\, ,\\
\label{free_equations_of_motion_mirror_model_massive_y}
&(\Box +1) y= 2 q \partial_0 y \, ,  &&(\Box +1) \bar{y}= -2 q \partial_0 \bar{y}.
\end{align}
\end{subequations}
If we define%
\footnote{Note that here and below we will write $p$ \textit{in lieu} of $\tilde{p}$
 for the mirror momentum to avoid burdening our notation. This should 
not (hopefully!) cause confusion because the momentum will be treated as
 a free variable. By contrast, for the dispersion we always use $\tilde{\omega}$ to highlight that it has a different functional form with respect to the original worldsheet~model.}
\begin{equation}
\lambda(p)= \sqrt{p^2+1 -q^2}
\end{equation}
then the dispersion relations of the massive and massless particles are 
\begin{equation}
\label{dispersion_relations_zy_and_barzbary_mirror}
\begin{split}
\tilde{\omega}_+(p)\equiv \tilde{\omega}_z(p)= \tilde{\omega}_y(p)= +i q +\lambda(p) \,,\\
\tilde{\omega}_-(p)\equiv \tilde{\omega}_{\bar{z}}(p)= \tilde{\omega}_{\bar{y}}(p)=-i q +\lambda(p) \, ,
\end{split}
\end{equation}
and
\begin{equation}
\label{dispersion_relation_masslss_in_mirror_model}
\tilde{\omega}_{\circ}(p)\equiv \tilde{\omega}_u(p)= \tilde{\omega}_v(p)=\tilde{\omega}_{\bar{u}}(p)= \tilde{\omega}_{\bar{v}}(p)=|p| \,,
\end{equation}
respectively. As expected the dispersion relations for the massless particles are left unchanged (note indeed that~\eqref{dispersion_relation_masslss_in_mirror_model} is equal to~\eqref{dispersion_relation_masslss_in_string_model}).

Differently from the string model, if $q \in \mathbb{R}$ the mirror theory is non-unitary; note indeed that if we take $q \in \mathbb{R}$ then both the energies in~\eqref{dispersion_relations_zy_and_barzbary_mirror} and the Lagrangian in~\eqref{Lagrangian_of_free_mirror_model} develop an imaginary part. The condition for the Lagrangian to be real and the energies of the mirror particles to be positive for all values of the momentum $p$ is that $q= i \tilde{q}$ with $\tilde{q} \in \mathbb{R}$.
This is the region of the $q$-parameter space where the mirror theory is unitary and we can compute amplitudes by applying standard perturbation theory around the vacuum where all the fields are equal to zero. This will be the approach that we will follow in the remaining part of this section. Finally, we may ask ourselves if it is possible to analytically continue these amplitudes to the values of $q$ where the theory is apparently non-unitary.
\begin{figure}
\begin{center}
    \begin{tikzpicture}
\tikzmath{\y=3;}

\filldraw[black] (0.43*\y,-1*\y)  node[anchor=west] {\scriptsize{$\bullet$}};
\draw[] (0.2*\y,-1.3*\y) -- (0.8*\y,-0.7*\y);
\draw[] (0.8*\y,-1.3*\y) -- (0.2*\y,-0.7*\y);
\draw[->] (0.1*\y,-1.3*\y) -- (0.3*\y,-1.1*\y);
\draw[->] (0.8*\y,-0.6*\y) -- (0.6*\y,-0.8*\y);
\draw[->] (0.2*\y,-0.6*\y) -- (0.4*\y,-0.8*\y);
\draw[->] (0.9*\y,-1.3*\y) -- (0.7*\y,-1.1*\y);
\filldraw[black] (-0.1*\y,-1.2*\y)  node[anchor=west] {\scriptsize{$X(p)$}};
\filldraw[black] (0.82*\y,-1.2*\y)  node[anchor=west] {\scriptsize{$X(p')$}};
\filldraw[black] (0.7*\y,-0.55*\y)  node[anchor=west] {\scriptsize{$\bar{X}(k)$}};
\filldraw[black] (0.1*\y,-0.55*\y)  node[anchor=west] {\scriptsize{$\bar{X}(k')$}};

\filldraw[black] (1.1*\y,-1.4*\y)  node[anchor=west] {\small{$\begin{aligned}
=& \mathscr{S}_X i \Bigl(-2   p'_0  k'_0  -2 p'_0 k_0 - 2 p_0 k'_0  - 2 p_0  k_0\\
&- i q \bigl[ (k_0+k_0') (p_0 p'_0-p_1 p_1' -1)- (p_0+p_0') (k_0 k_0'-k_1 k'_1 -1) \bigl] \Bigl)\\
&+i (2a -1) \Bigl( 2-2 (p_0 p_0'-p_1 p_1') (k_0 k_0'-k_1 k_1')\\
&-i q \bigl[ (k_0+k_0') (p_0 p_0'-p_1 p'_1 +1)- (p_0+p_0') (k_0 k_0'-k_1 k'_1 +1) \bigl] \Bigl)
\end{aligned}$}};


\filldraw[black] (0.43*\y,-1*\y-1.6*\y)  node[anchor=west] {\scriptsize{$\bullet$}};
\draw[] (0.2*\y,-1.3*\y-1.6*\y) -- (0.8*\y,-0.7*\y-1.6*\y);
\draw[] (0.8*\y,-1.3*\y-1.6*\y) -- (0.2*\y,-0.7*\y-1.6*\y);
\draw[->] (0.1*\y,-1.3*\y-1.6*\y) -- (0.3*\y,-1.1*\y-1.6*\y);
\draw[->] (0.8*\y,-0.6*\y-1.6*\y) -- (0.6*\y,-0.8*\y-1.6*\y);
\draw[->] (0.2*\y,-0.6*\y-1.6*\y) -- (0.4*\y,-0.8*\y-1.6*\y);
\draw[->] (0.9*\y,-1.3*\y-1.6*\y) -- (0.7*\y,-1.1*\y-1.6*\y);
\filldraw[black] (-0.1*\y,-1.2*\y-1.6*\y)  node[anchor=west] {\scriptsize{$y(p)$}};
\filldraw[black] (0.82*\y,-1.2*\y-1.6*\y)  node[anchor=west] {\scriptsize{$z(p')$}};
\filldraw[black] (0.7*\y,-0.55*\y-1.6*\y)  node[anchor=west] {\scriptsize{$\bar{y}(k)$}};
\filldraw[black] (0.1*\y,-0.55*\y-1.6*\y)  node[anchor=west] {\scriptsize{$\bar{z}(k')$}};

\filldraw[black] (1*\y,-1*\y-2.3*\y)  node[anchor=west] {\small{$\begin{aligned}
&=i \Bigl( -p_1 k_1 - p_0k_0 + p_1'k_1' + p_0'k_0'\\
&+\frac{i q}{2}\bigl[ (p_0'k_1'+k_0'p_1')(k_1-p_1)+ (p_0k_1 + k_0p_1)(p_1'-k_1')\\
&- (1-p_1'k_1'-p_0'k_0')(p_0-k_0)+ (1-p_1 k_1 -p_0 k_0)(p_0'-k_0')\bigl] \Bigl)\\
&+ i (2a-1) \Bigl( 1 + (p'_0 k'_1 +k'_0 p'_1) (p_0 k_1 +k_0 p_1) - (p'_1 k'_1 +p'_0 k'_0) (p_1 k_1 +p_0 k_0)\\
&-\frac{i q}{2} \bigl[ (k_0'-p_0') (-p_1 k_1-p_0 k_0+1) + (k_0-p_0) (-p'_1 k'_1-p'_0 k'_0+1)\\
&- (p_1-k_1) (p'_1 k'_0+p'_0 k'_1) - (p'_1-k'_1) (p_1 k_0+p_0 k_1) \bigl] \Bigl)
\end{aligned}$}};


\draw[] (0.2*\y,-1.3*\y-3.6*\y) -- (0.8*\y,-0.7*\y-3.6*\y);
\draw[] (0.8*\y,-1.3*\y-3.6*\y) -- (0.2*\y,-0.7*\y-3.6*\y);
\draw[->] (0.1*\y,-1.3*\y-3.6*\y) -- (0.3*\y,-1.1*\y-3.6*\y);
\draw[->] (0.8*\y,-0.6*\y-3.6*\y) -- (0.6*\y,-0.8*\y-3.6*\y);
\draw[->] (0.2*\y,-0.6*\y-3.6*\y) -- (0.4*\y,-0.8*\y-3.6*\y);
\draw[->] (0.9*\y,-1.3*\y-3.6*\y) -- (0.7*\y,-1.1*\y-3.6*\y);
\filldraw[black] (-0.1*\y,-3.6*\y-1.2*\y)  node[anchor=west] {\scriptsize{$\mu(p)$}};
\filldraw[black] (0.82*\y,-1.2*\y-3.6*\y)  node[anchor=west] {\scriptsize{$X(p')$}};
\filldraw[black] (0.7*\y,-0.55*\y-3.6*\y)  node[anchor=west] {\scriptsize{$\bar{\mu}(k)$}};
\filldraw[black] (0.1*\y,-0.55*\y-3.6*\y)  node[anchor=west] {\scriptsize{$\bar{X}(k')$}};

\filldraw[black] (0.43*\y,-1*\y-3.6*\y)  node[anchor=west] {\scriptsize{$\bullet$}};

\filldraw[black] (1.1*\y,-1*\y-4*\y)  node[anchor=west] {\small{$\begin{aligned}
=& \mathscr{S}_X i \Bigl( - (p_1k_1+p_0k_0)\\
&+\frac{iq}{2} \bigl[ (p_0k_1 + k_0p_1)(p_1'-k_1') - (p_1k_1+p_0 k_0)(p_0'-k'_0) \bigl] \Bigl)\\
&+i (2a-1) \Bigl( (p_0' k_1' + p_1' k_0')(p_0 k_1 + p_1 k_0)-(p_1' k_1' + p_0' k_0')(p_1 k_1 + p_0 k_0)\\
&+\frac{i q}{2} \bigl[ (p_1 k_0 + p_0 k_1) (p_1'-k_1') - (p_1 k_1 + p_0 k_0) (p_0'-k_0') \bigl]\Bigl)
\end{aligned}$}};

\end{tikzpicture}
\end{center}
\caption{Massive and massive-massless interaction vertices of the bosonic mirror model. 
The sign function~\eqref{eq:sign_X_z_y} has been used in the first and last vertex.}
\label{image_vertex_interactions_mirror_theory}
\end{figure}

The canonical quantization conditions~\eqref{canonical_quantisation_conditions_for_z_string}, together with the form of the conjugate momenta
\begin{equation}
P_z = \frac{\partial \mathcal{L}_m^{(2)}}{\partial \dot{z}}=\dot{\bar{z}} +q \bar{z}\hspace{3mm},\hspace{3mm} P_{\bar{z}} = \frac{\partial \mathcal{L}_m^{(2)}}{\partial \dot{\bar{z}}}=\dot{z} -qz \,,
\end{equation}
require to expand the massive fields as follows
\begin{equation}
\label{z_and_barz_in_terms_of_creation_and_annihilation_operators_mirror}
\begin{aligned}
&z(\tau,\sigma) = \int \frac{dp}{\sqrt{(2\pi)} \sqrt{2 \lambda(p)}}\biggl[e^{-i(\tilde{\omega}_+(p)\tau - p\sigma) }a_z(p) + e^{i(\tilde{\omega}_-(p)\tau - p\sigma)} a^\dagger
_{\Bar{z}}(p)\biggr]\, ,\\
&\Bar{z}(\tau,\sigma) = \int \frac{dp}{\sqrt{(2\pi)} \sqrt{2 \lambda(p)}}\biggl[e^{-i(\tilde{\omega}_-(p)\tau - p\sigma) }a_{\Bar{z}}(p) + e^{i(\tilde{\omega}_+(p)\tau - p\sigma) }a^\dagger
_z(p)\biggr]\, .
\end{aligned}
\end{equation}
The conditions in~\eqref{canonical_quantisation_conditions_for_z_string} are then satisfied if the creation and annihilation operators satisfy the commutation relations in~\eqref{commutation_relations_between_a_and_dagger_a}. The expansion of $y$ and $\bar{y}$ in terms of creation and annihilation operator is the same as the expansion of $z$ and $\bar{z}$.
It is worth stressing that the decomposition in oscillators is formally possible for any~$q\in\mathbb{C}$, without affecting the commutation relations.

\subsection{Tree-level scattering in the mirror model}
\label{Section_on_tree_level_S_matrix_elements_of_the_mirror_model}

Similarly to what we did for the free Lagrangian, the interaction Lagrangian of the mirror model is obtained by performing the substitution~\eqref{eq:map_string_mirror_derivatives} in the interaction Lagrangian of the string model. 
In appendix~\ref{appendix:Lagrangian_mirror_model} we provide the mirror Lagrangian to the fourth-order in the fields expansion.
From such a Lagrangian the four-point interaction vertices can be computed: we summarise the massive and massive-massless vertices in figure~\ref{image_vertex_interactions_mirror_theory} where we assume all the particles to be incoming. The massless vertices are instead left invariant under the mirror transformation and were already shown in figure~\ref{image_vertex_interactions_string_theory_massless}.
From the way in which energies and momenta of the string and mirror particles are connected (see~\eqref{eq:string_mirror_energies_and_momenta}) the vertices of the mirror model can easily be generated from the vertices of the string model by replacing $p_0 \to i \tilde{p}_1$ and  $p_1 \to i \tilde{p}_0$.
Once the vertices are known, the tree-level S-matrix of the mirror model can be computed: this is done by putting the external particles on-shell in the vertices and multiplying by the following additional factors
\begin{equation}
\label{extra_multiplicative_factors_to_be_added_in_mirror_amplitudes}
\begin{split}
&\text{for each external massive particle:} \hspace{13mm} \frac{1}{\sqrt{2 \lambda}}  \ ;\\
&\text{for each external massless particle:} \hspace{12mm} \frac{1}{\sqrt{2 \tilde{\omega}_\circ}}  \ ;\\
&\text{an overall multiplicative factor:} \hspace{6mm} \delta(\sum_{i} p^{\text{in}}_i -\sum_{i} p^{\text{out}}_i) \delta(\sum_{i} \tilde{\omega}^{\text{in}}_i -\sum_{i} \tilde{\omega}^{\text{out}}_i) .
\end{split}\end{equation}

As for the string model, the mirror theory has different symmetries which connect the tree-level S-matrix elements to each other; for the mirror theory these symmetries are \textit{braiding unitarity}, \textit{crossing} (already summarised in equation~\eqref{eq:braiding_unit_crossing_string}) and \textit{parity}. This last symmetry is due to the fact that the terms composing the mirror Lagrangian summarised in section~\ref{appendix:Lagrangian_mirror_model} are invariant under a transformation $\sigma \to - \sigma$. 
At the level of the S-matrix, we obtain
\begin{equation}
\label{P_symmetry_on_tree_mirror_amplitudes}
\text{parity:} \ \ \ \tr_{AB}(\{p_0, p_1\}, \{q_0, q_1\})=\tr_{B A}(\{q_0, -q_1\}, \{p_0, -p_1\}) \,.
\end{equation}
 
Below we summarise the S-matrix elements of the mirror model, using the CDD factor $\Phi_{ij}(p,p')$ of eq.~\eqref{eq:CDD_TbarT}, which now depends on the mirror dispersions $\tilde{\omega}_{+}$, $\tilde{\omega}_{-}$ or $\tilde{\omega}_{\circ}$.
\paragraph{Massive-massive}
\begin{equation}
\label{eq:massive_massive_mirror_S_matrices}
\begin{aligned}
& \mathbb{T}\ket{X_{\pm}(p)X_{\pm}(p')} = \Bigl( \mathscr{S}_X  \frac{(\tilde{\omega}_{\pm}+\tilde{\omega}'_{\pm})(\tilde{\omega}_{\pm} \, p'+\tilde{\omega}'_{\pm} \, p_)}{2(\tilde{\omega}_{\pm} - \tilde{\omega}'_{\pm} )}  +\Phi_{\pm\pm}(p, p') \Bigl)\ket{X_{\pm}(p')X_{\pm}(p)} ,\\
& \mathbb{T}\ket{X_{\pm}(p)X_{\mp}(p')} = \Bigl( \mathscr{S}_X  \frac{(\tilde{\omega}_{\pm}-\tilde{\omega}_{\mp}')(\tilde{\omega}_{\pm} \, p'+\tilde{\omega}_{\mp}' \, p)}{2(\tilde{\omega}_{\pm} +\tilde{\omega}_{\mp}')}  +\Phi_{\pm\mp}(p, p') \Bigl)\ket{X_{\mp}(p')X_{\pm}(p)} ,\\
&\mathbb{T}\ket{z_{\pm}(p)y_{\pm}(p')} = \Bigl( -\frac{1}{2}(\tilde{\omega}_{\pm} \, p' +\tilde{\omega}_{\pm}' \, p )  + \Phi_{\pm\pm}(p, p') \Bigl) \ket{y_{\pm}(p') z_{\pm}(p)} ,\\
&\mathbb{T}\ket{z_{\pm}(p)y_{\mp}(p')} = \Bigl( -\frac{1}{2}(\tilde{\omega}_{\pm} \, p' +\tilde{\omega}_{\mp}' \, p )  + \Phi_{\pm\mp}(p, p') \Bigl) \ket{y_{\mp}(p') z_{\pm}(p)} .
\end{aligned}
\end{equation}
\paragraph{Mixed-mass}
\begin{equation}
\label{eq:massive_massless_mirror_S_matrices}
\begin{aligned}
&\mathbb{T}\ket{X_{\pm}(p) U (p')} = \Bigl( -\mathscr{S}_X \frac{1}{2}(\tilde{\omega}_{\pm} \, p'+\tilde{\omega}_{\circ}' \, p)+ \Phi_{\pm\circ}(p, p')\Bigl) \ket{ U(p') X_{\pm}(p)} ,\\
&\mathbb{T}\ket{U(p) X_{\pm} (p')} = \Bigl( \mathscr{S}_X \frac{1}{2}(\tilde{\omega}_{\circ} \, p'+\tilde{\omega}_{\pm}' \, p)+ \Phi_{\circ\pm}(p, p') \ket{ X_{\pm}(p') U(p)} .
\end{aligned}
\end{equation}

\paragraph{Massless}
\begin{equation}
\label{eq:massless_massless_mirror_S_matrices}
\mathbb{T} \ket{U(p) V(p')} = \Phi_{\circ\circ}(p, p') \ket{V(p') U(p)} \,.
\end{equation}
Differently from the string model the requirement $v>0>v'$ in the mirror theory translates into the condition $p>0>p'$ on the momenta. 
The conventions for the fields and dispersion relations are the same already used in the string model. The only difference is the functional dependence of the dispersion relations on the momenta, which are reported in~\eqref{dispersion_relations_zy_and_barzbary_mirror} and~\eqref{dispersion_relation_masslss_in_mirror_model}.

As expected, we observe that the tree-level S-matrix elements of the mirror model can be obtained from the S-matrix elements of the string model by replacing $p \to i \tilde{\omega}$ and $\omega \to i \tilde{p}$.
We will see that this guarantees a match with the expansion of the all-loop result~\cite{Lloyd:2014bsa}.

\subsection{Comparison with non-perturbative results}
\label{sec:comparison}

In this section, we briefly check that the S~matrices found so far agree with the non-perturbative results for the mixed flux S-matrix found in the literature
at the leading order in $\frac{1}{T}$.

Following the conventions in~\cite{Eden:2021xhe} the energies of the different string particles are given by
\begin{equation}
\label{eq:exact_dispersion_relation}
E(p,m)= \sqrt{\bigl( m+ \frac{k}{2 \pi} p \bigl)^2 +4 \, h^2 \sin^2 \bigl( \frac{p}{2} \bigl)}
\end{equation}
where $k$ is the quantised amount of NS-NS flux and is defined as
\begin{equation}
\frac{k}{2 \pi} = q \, T \,.
\end{equation}
The parameter $h$ is instead a function of $q$ and of the string tension which for $T \gg 1$ becomes~\cite{Lloyd:2014bsa}
\begin{equation}
h= T \sqrt{1 - q^2} \,.
\end{equation}
Scaling $p \to \frac{p}{T}$ and taking the limit in which $T \gg 1$ we obtain
\begin{equation}
\label{eq:limit_string_energy}
E(p,m) \to \omega(p,m)= \sqrt{p^2 + m^2 +2 m q p} \,.
\end{equation}
Comparing this expression with the dispersion relations in~\eqref{dispersion_relations_zy_and_barzbary} we note that
the fundamental particles $z$ and $y$ have $m=+1$, while the particles $\bar{z}$ and $\bar{y}$ have $m=-1$. Massless particles have instead $m=0$.
The exact S-matrix elements associated with the scattering of these particles (up to certain unknown dressing factors) are summarised in~\cite{Eden:2021xhe} in terms of the Zhukovsky variables
\begin{equation}
\label{eq:Zhukovsky}
\begin{split}
x_L^{\pm}&= \frac{ 1+ \frac{k}{2 \pi} p + E(p,+1)} {2h \sin \bigl( \frac{p}{2} \bigl)}\,e^{\pm i\frac{p}{2}}\,,\\
x_R^{\pm}&= \frac{ 1- \frac{k}{2 \pi} p + E(p,-1)}{2h \sin \bigl( \frac{p}{2} \bigl)}\, e^{\pm i\frac{p}{2}}\,.
\end{split}
\end{equation}
Applying the scaling $p \to \frac{p}{T}$ and taking the limit of large string tension we obtain 
\begin{equation}
\label{eq:Limit_Zhuk_variables}
\begin{split}
x_L^{\pm}&\to \frac{ 1+ q p + \sqrt{ 1+ 2q p + p^2 }}{p \sqrt{1-q^2}} \bigl(1 \pm i \frac{p}{2T} \bigl)\,,\\
x_R^{\pm}&\to \frac{ 1- q p + \sqrt{ 1- 2q p + p^2 }}{p \sqrt{1-q^2}} \bigl(1 \pm i \frac{p}{2T} \bigl)\,.
\end{split}
\end{equation}
Following the standard notation used in literature the labels $L$ and $R$ stand for ``left'' and ``right'' representations, associated with $m=+1$ and $m=-1$ respectively (the origin of the name is that the corresponding excitations should be related to left- and right-movers in the dual CFT, see e.g.~\cite{Frolov:2023pjw}) for a recent discussion.

The integrable bootstrap~\cite{Lloyd:2014bsa} fixes the S-matrix elements for the scattering of particles in the different massive sectors (which are left-left, left-right, right-left and right-right) up to four overall dressing factors; those in turn have to be fixed by more subtle consideration such as crossing and analyticity (the same argument applies to the massless excitations).
To check the matching of the S~matrix up to the dressing factors we can consider the $1/T$ expansion of ratios of the type
\begin{equation}
\label{eq:comparison_log_ratio_S}
-i \log \frac{\sr_{AB}(p, p')}{\sr_{CD}(p, p')} = \frac{1}{T}\left(\tr_{AB}(p, p')- \tr_{CD}(p, p')\right) +O(T^{-2})\,.
\end{equation}
For the string model, the S-matrix elements on the LHS of the equality can be taken from~\cite{Eden:2021xhe} and expanded using~\eqref{eq:Limit_Zhuk_variables}, while the quantities $\tr_{ij}(p, p')$ on the RHS of~\eqref{eq:comparison_log_ratio_S} are the tree-level interaction elements summarised in equations~\eqref{eq:massive_massive_string_S_matrices}, \eqref{eq:massive_massless_string_S_matrices} and~\eqref{eq:massless_massless_string_S_matrices}.
As expected all Feynman diagrams contributing at the order $T^{-1}$ are purely transmitting when we restrict to bosonic processes. Reflection elements are also contained in the all-loop S~matrix, but they appear at order~$T^{-2}$ (\textit{i.e.} at one loop).%
\footnote{For instance for left-left scattering, in the notation of~\cite{Eden:2021xhe} the transmission elements are $A^2_{LL}$, $F^2_{LL}$, $B^2_{LL}$ and $D^2_{LL}$; the bosonic reflection elements are instead $C^2_{LL}$ and $E^2_{LL}$, and the boson-boson-fermion-fermion processes are of the from $C_{LL}B_{LL}$, $E_{LL}D_{LL}$, \textit{etc.}. At $T\gg1$ we find that  $A_{LL}$,$B_{LL}$,$D_{LL}$, $F_{LL}$ go as $\pm1+\mathcal{O}(T^{-1})$, while  $C_{LL}$ and $E_{LL}$ are of order $\mathcal{O}(T^{-1})$.}
The value of an individual perturbative S-matrix element, such as~$\tr_{AB}(p, p')$, provides a test for the dressing factor (of course for the string model these results were already found at the tree level and beyond in the literature, see e.g.~\cite{Sundin:2016gqe} and references therein). Recently, the massive dressing factors for the string model were proposed in~\cite{Frolov:2024pkz} and the authors found agreement with the perturbative massive S~matrices (the dressing factors for the massless processes have not yet been studied).

In the case of the mirror model instead, the dressing factors have not yet been discussed in the literature; we expect this to happen in the near future, as in principle they should follow by analytic continuation of the results of~\cite{Frolov:2024pkz}. Our results for the normalisation of the S~matrix are therefore a (new) constraint on the dressing factors.
It is more straightforward to compare our results with the integrable bootstrap construction of the mirror model \textit{up to dressing factors}, i.e.~in terms of ratios of the form~\eqref{eq:comparison_log_ratio_S}. The mirror-model S-matrix representations can be obtained by the string-model ones by analytic continuation along the lines of~\cite{Frolov:2021zyc}.
This continuation is straightforward in terms of the Zhukovsky variables~$x^\pm_{L,R}$ and, since the ratios~\eqref{eq:comparison_log_ratio_S} are rational functions of the Zhukovsky variables, it can be done by literally replacing the ``string'' Zhukovsky variables by their ``mirror'' counterparts, where energy and momentum have undergone the double Wick rotation~\eqref{eq:string_mirror_energies_and_momenta}.%
\footnote{This is not so for the dressing factors, which have an intricate cut structure when expressed  terms of $x^{\pm}_{L,R}$, see~\cite{Frolov:2024pkz}.}
Performing the double Wick rotation on~\eqref{eq:exact_dispersion_relation} we find the mirror dispersion relation,
\begin{equation}
\label{eq:exact_mirror_dispersion_relations}
4 h^2 \sinh^2 \bigl( \frac{\tilde{\omega}}{2} \bigl) - \left( m +i \frac{k}{2\pi} \tilde{\omega} \right)^2 - \tilde{p}^2=0 \,.
\end{equation}
While it is not possible to invert~\eqref{eq:exact_mirror_dispersion_relations} and write the mirror dispersion $\tilde{\omega}$ in terms of elementary functions, it is immediate to see that for $\tilde{p} \in \mathbb{R}$ and $m\neq0$,  $\tilde{\omega}$ picks up an imaginary part. The mirror Zhukovsky variables can then be written by performing the same double Wick rotation in~\eqref{eq:Zhukovsky}, and this gives the integrability prediction for the exact S~matrix (or at least, for its matrix part).

By applying the same type of expansion as we did for the string model, which in this case corresponds to replace $\tilde{\omega} \to \tilde{\omega}/T$ and take $T \gg 1$, we can work out the dispersion and S-matrix order by order in~$1/T$. At leading order the mirror dispersion can be simply written as
\begin{equation}
\label{eq:limit_mirror_energy}
\tilde{\omega}(m, \tilde{p}) = i m q + \sqrt{\tilde{p}^2 + m^2 (1-q^2)} \,.
\end{equation}
As one may have expected, this  is exactly the dispersion relation for the left ($m=+1$) and right ($m=-1$) particles obtained in~\eqref{dispersion_relations_zy_and_barzbary_mirror} from the Lagrangian expansion.
It is then straightfoward to check that this expansion matches our tree-level computation by comparing it with~\eqref{eq:massive_massive_mirror_S_matrices}, \eqref{eq:massive_massless_mirror_S_matrices} and~\eqref{eq:massless_massless_mirror_S_matrices}.

\paragraph{Pseudo-unitarity (or lack thereof).}
We have stressed that for real values of $q$ the mirror S~matrix is not unitary. There are however weaker notions of unitarity, such as pseudo-unitarity, which is
\begin{equation}
\label{eq:pseudounitairty_property}
S^{\dagger}(p_1, p_2)= B\, S^{-1}(p^*_1, p^*_2)\, B^{-1} \,,
\end{equation}
where $B$ is a hermitian matrix.
This condition has appeared in the study of worldsheet S-matrices before, namely in the context of deformations of AdS$_5 \times$ S$^5$~\cite{Arutyunov:2012zt}, and in all known models in which this property is satisfied it turned out that the spectrum was self-conjugate or real\footnote{This condition comes from the study of pseudo-hermitian Hamiltonians, i.e. Hamiltonians satisfying the condition
$$
H^\dagger=B H B^{-1}
$$
with $B$ hermitian. It is a known fact that all matrices $H$ with real spectrum satisfy this property~\cite{Mostafazadeh:2001jk}. We remand to appendix~A of~\cite{Arutyunov:2012zt} for a more detailed discussion.}. 
It is interesting to note that this condition is indeed satisfied by the tree-level mirror S~matrix.
It appears however that this is not true at higher loops: if we take the S-matrix to be given (up to a normalisation) by the all-loop integrable S~matrix~\cite{Lloyd:2014bsa}, continued to the mirror region with the requirement that the Zhukovsky variables transform as in~\eqref{eq:mirror_Zh_transormation}, we can try to solve~\eqref{eq:pseudounitairty_property} as a linear equation for~$B$. It appears that this has no solution. However, we shall see below that the S~matrix nonetheless satisfies conjugacy properties sufficient to ensure reality of wrapping corrections. 

\subsection{Mirror-string interactions and L\"uscher corrections}
\label{sec:Luscher_corrections}

We observed that the S-matrix elements of the mirror model can be obtained from the S-matrix elements of the string model by applying the map $p \to i \tilde{\omega}$ and $\omega \to i \tilde{p}$. It is therefore natural to generate an S-matrix describing the interactions between mirror and string particles by applying the mirror transformation on only one particle.
Below we summarise the S-matrix elements describing the interaction of a mirror particle coming from the left with a string particle coming from the right. We define
\begin{equation}
\mathcal{F}_{ij}(\tilde{p}, p')= \frac{2 a -1}{2} \bigl[ \tilde{p} \, p' - \tilde{\omega}_i(\tilde{p}) \omega_j(p') \bigl]
\end{equation}
and use a superscript index $m$/$s$ to identify the mirror/string particles.

\paragraph{Massive-massive}

\begin{equation}
\label{eq:massive_massive_S_matrix_mirror_string}
\begin{aligned}
& \mathbb{T}\ket{X_{\pm}^{(m)}(\tilde{p})X_{\pm}^{(s)}(p')} = i \Bigl( -\mathscr{S}_X  \frac{(i \tilde{\omega}_{\pm}+p')( \tilde{\omega}_{\pm} \omega'_{\pm} + \tilde{p} \, p')}{2(i \tilde{\omega}_{\pm} -p')}  +\mathcal{F}_{\pm \pm}(\tilde{p}, p') \Bigl)\ket{X^{(s)}(p')X^{(m)}(\tilde{p})} \,,\\
& \mathbb{T}\ket{X_{\pm}^{(m)}(\tilde{p})X_{\mp}^{(s)}(p')} = i \Bigl( -\mathscr{S}_X  \frac{(i \tilde{\omega}_{\pm}-p')( \tilde{\omega}_{\pm} \omega_{\mp}' + \tilde{p} \, p')}{2(i \tilde{\omega}_{\pm} +p')}  +\mathcal{F}_{\pm \mp}(\tilde{p}, p') \Bigl)\ket{X_{\mp}^{(s)}(p')X_{\pm}^{(m)}(\tilde{p})} \,,\\
&\mathbb{T}\ket{z_{\pm}^{(m)}(\tilde{p})y_{\pm}^{(s)}(p')} = i \Bigl( \frac{1}{2}(   \tilde{\omega}_{\pm} \, \omega_{\pm}'+ \tilde{p} \, p' )  + \mathcal{F}_{\pm \pm}(\tilde{p}, p') \Bigl) \ket{y^{(s)}_{\pm}(p') z^{(m)}_{\pm}(\tilde{p})} \,,\\
&\mathbb{T}\ket{z_{\pm}^{(m)}(\tilde{p})y_{\mp}^{(s)}(p')} = i \Bigl( \frac{1}{2}(   \tilde{\omega}_{\pm} \, \omega_{\mp}'+ \tilde{p} \, p' )  + \mathcal{F}_{\pm \mp}(\tilde{p}, p') \Bigl) \ket{y^{(s)}_{\mp}(p') z^{(m)}_{\pm}(\tilde{p})} \,.
\end{aligned}
\end{equation}
\paragraph{Mixed-mass}
\begin{equation}
\label{eq:massive_massless_S_matrix_mirror_string}
\begin{aligned}
&\mathbb{T}\ket{X_{\pm}^{(m)}(\tilde{p}) U^{(s)} (p')} = i \Bigl( \mathscr{S}_X \frac{1}{2}(\tilde{p} \, p'+ \tilde{\omega}_{\pm} \, \omega_{\circ}')+ \mathcal{F}_{\pm \circ}(\tilde{p}, p') \Bigl) \ket{ U^{(s)}(p') X_{\pm}^{(m)}(\tilde{p})} \,,\\
&\mathbb{T}\ket{U^{(m)}(\tilde{p}) X_{\pm}^{(s)} (p')} = i \Bigl( -\mathscr{S}_X \frac{1}{2}(\tilde{p} \, p'+ \tilde{\omega}_\circ \, \omega_{\pm}')+ \mathcal{F}_{\circ \pm}(\tilde{p}, p') \Bigl) \ket{ X_{\pm}^{(s)}(p') U^{(m)}(\tilde{p})} \,.
\end{aligned}
\end{equation}
\paragraph{Massless}
\begin{equation}
\label{eq:massless_massless_S_matrix_mirror_string}
\mathbb{T} \ket{U^{(m)}(\tilde{p}) V^{(s)}(p')} = i \mathcal{F}_{\circ \circ}(\tilde{p}, p') \ket{V^{(s)}(p') U^{(m)}(\tilde{p})} \,,\hspace{5mm} U,V= u, v, \bar{u}, \bar{v} \,.
\end{equation}
These S-matrix elements are important to study finite volume corrections to the quantised energy levels of the string as discussed below.

The energy levels of an integrable quantum field theory defined on a circle of very large length $\stl$ can be approximately described in terms of the asymptotic Bethe Ansatz Equations. These will give real energies because, in the worldsheet theory, the dispersion is real and the S~matrix is unitary.  
However, for $\stl$ finite the energy levels get modified by the so-called L\"uscher corrections (see~\cite{Janik:2010kd} for a review): these corrections are generated by mirror particles travelling around the circle and scattering once at a time with all the physical (or string) particles on the circle~\cite{Luscher:1985dn}. When the mirror particles are massive, such effects are exponentially suppressed as~$e^{-\tilde{\omega}\stl}$, while for gapless mirror particles they are expected to affect the energy at order~$1/\stl$~\cite{Brollo:2023rgp}.
The L\"uscher corrections do not fully account for all finite-size effects, but it is instructive to study whether they are real.

Let us consider a state containing $N$ worldsheet excitations of types $\{a_1, \dots, a_N\}$ carrying momenta $\{p_1, \dots, p_N\}$. The momenta will have to satisfy the asymptotic Bethe equations of the \textit{original} (as opposed to \textit{mirror}) model. In particular, they will have to be real or come in complex-conjugate complexes to ensure the reality of the energy.
The correction to the energy due to the ``F-term''%
\footnote{%
The F-term is related to the wrapping of virtual particles around the cylinder; an additional contribution is due to the $\mu$-term, which has to do with bound-states splitting and recombining across the cylinder~\cite{Janik:2010kd}; we will not discuss this term here because we do not have control over bound states at the tree level.} 
takes a simple form~\cite{Bajnok:2008bm}:
\begin{equation}
\label{eq:Luscher_correction}
\Delta E = -\sum_{Q} \int_{-\infty}^{+\infty} \frac{d \tilde{p}}{2 \pi} e^{-\tilde{\omega}_Q (\tilde{p})\stl}  \Lambda_{Q; a_1,\dots a_N} (\tilde{p}; p_1,\dots p_N) \,,
\end{equation}
where the sum runs over all the types of mirror particles~$Q$ (more properly, over all mirror representations) travelling around the circle with energy $\tilde{\omega}_Q$ and momentum $\tilde{p}$.
$\Lambda(\tilde{p};p_1,\dots p_N)$ is the eigenvalue of the transfer matrix related to the scattering with particles of flavour $a_1,\dots a_N$ and string momentum~$p_1,\dots p_N$.
These corrections should correspond to the leading order expansion of the TBA equations at large $\stl$.

If we expand these corrections in~$1/T$,
\begin{equation}
\Delta E =\Delta E^{(0)}+ \frac{1}{T} \Delta E^{(1)}+ \frac{1}{T^2} \Delta E^{(2)} + \dots \,,
\end{equation}
where $T$ is the usual string tension, then $\Delta E^{(0)}$ and $\Delta E^{(1)}$ are completely captured by the tree-level S-matrix and can be written as%
\footnote{The factor of two is due to the fact that there are two bosons in each representation; for instance, $z$ and $y$ have $Q=1$.}
\begin{equation}
\label{eq:E0_correction}
\Delta E^{(0)} =  -2\sum_{Q} \int_{-\infty}^{+\infty} \frac{d \tilde{p}}{2 \pi} e^{-\tilde{\omega}_Q(\tilde{p}) \stl} 
\end{equation}
and
\begin{equation}
\label{eq:E1_correction}
\Delta E^{(1)} = -i \sum_{Q}\sum_{j\in Q} \sum_{n=1}^N \int_{-\infty}^{+\infty} \frac{d \tilde{p}}{2 \pi} e^{-\tilde{\omega}_Q(\tilde{p}) \stl}   \tr_{j a_n} (\tilde{p}, p_n) \,.
\end{equation}
The scalar factors $\tr_{j a_n}$ in~\eqref{eq:E1_correction} are listed in
equations~\eqref{eq:massive_massive_S_matrix_mirror_string}, \eqref{eq:massive_massless_S_matrix_mirror_string} and~\eqref{eq:massless_massless_S_matrix_mirror_string}.

The expression in~\eqref{eq:E0_correction} is easily computed using the dispersion relations for the different mirror particles; the result can be split into a massive (carried by $j=z,\,y,\,\bar{z},\, \bar{y}$ in the sum) and a massless (carried by $j=u,\,v,\,\bar{u},\, \bar{v}$) contribution, leading to
\begin{equation}
\label{eq:E0_correction_computed}
\begin{split}
\Delta E^{(0)} &=  \frac{2}{\pi} \cos{(q \stl)} \int_{-\infty}^{+\infty} d \tilde{p} e^{-\lambda(\tilde{p}) \stl} + \frac{2}{\pi} \int_{-\infty}^{+\infty} d \tilde{p} e^{-|\tilde{p}| \stl}\,,
\end{split}
\end{equation}
which is perfectly real and convergent for all values of $q \in [0,1]$. This is a remarkable fact since the dispersion relations of the mirror massive particles are complex and the mirror theory is non-unitary, and it is due to the fact that particles come in pairs --- a consequence of charge-conjugation invariance.
A few remarks are in order. Firstly, we actually expect this model to have bound states both in the string and mirror model~\cite{Frolov:2023lwd}. Their tree-level dispersion would still be given by
\begin{equation}
    \tilde{\omega}(\tilde{p},m)=\sqrt{\tilde{p}^2+(1-q^2)\,m^2}+i\,q\,m\,,
\end{equation}
but with $m=\pm2, \pm3, \dots$, up to infinity. Clearly, this does not affect our argument but just extends the sum over~$Q$ to infinitely many values. It is easy to see that the sum is convergent, as the mirror-energy contribution is exponentially suppressed for $|m|\to\infty$. Finally, while this argument nicely illustrates how the energy can remain real, at this order it is not necessary had we considered the full theory. In fact, due to the presence of the fermions (and supersymmetry), this momentum-independent term would be immediately zero there.

Reality persists also for the contribution in~\eqref{eq:E1_correction} (which is in general non-zero) also in the full theory. Indeed from the mirror dispersion relations in section~\ref{subsec_mirror_theory} and the S-matrix elements in~\eqref{eq:massive_massive_S_matrix_mirror_string}, \eqref{eq:massive_massless_S_matrix_mirror_string} and~\eqref{eq:massless_massless_S_matrix_mirror_string}, we note that for all values of $q \in [0,1]$ it holds that, for real momenta,
\begin{equation}
\label{eq:Tmatrixconj}
\tilde{\omega}_{-Q}=\big(\tilde{\omega}_{Q}\big)^* \hspace{4mm} \text{and} \hspace{4mm}\tr_{\bar{\jmath} a} (\tilde{p}, p)=-\big(\tr_{j a} (\tilde{p}, p)\big)^* \,,
\end{equation}
for any mirror particle $j$ in the representation~$Q$ (conjugate to a particle~$\bar{\jmath}$ in representation~$-Q$) and string particle $a$. The expression in~\eqref{eq:E1_correction} can then be written as
\begin{equation}
\begin{split}
\Delta E^{(1)} =& + \sum_{Q\geq 1}\sum_{j\in Q} \sum_{n=1}^N \int_{-\infty}^{+\infty} \frac{d \tilde{p}}{\pi} \text{Im} \bigl( e^{-\tilde{\omega}_Q(\tilde{p}) \stl}   \tr_{j a_n} (\tilde{p}, p_n) \bigl)\\
&+\sum_{j\in \{u, v\} } \sum_{n=1}^N \int_{-\infty}^{+\infty} \frac{d \tilde{p}}{\pi} \text{Im} \bigl( e^{-\tilde{\omega}_0(\tilde{p}) \stl}   \tr_{j a_n} (\tilde{p}, p_n) \bigl)
\end{split}
\end{equation}
where the second row takes into account the massless representation, with associated conjugate representation composed of $\{\bar{u}, \bar{v}\}$.

We conclude that even though the mirror theory has a complex Hamiltonian, at the leading order in $T^{-1}$ the quantised energy levels of the string are real. This argument assumes the presence of mirror particles of complex conjugate energies, which remains true at all~loops. As for the structure scattering, we need to investigate the properties of the transfer matrix under conjugation, see~\eqref{eq:Luscher_correction}. The transfer matrix for the string and mirror model was computed in~\cite{Seibold:2022mgg}. That was done for the pure-RR case, but at least for the matrix part of the transfer matrix we can write similar formulae in terms of modified Zhukovsky variable. It is easy to see, as far as the matrix structure of the S~matrix is concerned, a sufficient condition for the L\"uscher F-term to be real is that the mirror Zhukovsky variables obey
\begin{equation}
\label{eq:mirror_Zh_transormation}
    \big(\tilde{x}^\pm(\tilde{p},m)\big)^*= \frac{1}{\tilde{x}^{\mp}(\tilde{p},-m)}\,,
\end{equation}
for $\tilde{p}$ real.
Indeed, this is the reality condition of the mirror theory~\cite{Frolov:2023lwd}.
A conjugacy condition similar to~\eqref{eq:Tmatrixconj}  may be derived for the transmission processes.
In this case we should work in terms of the eigenvalues of the full transfer matrix~\eqref{eq:Luscher_correction}. The eigenvalues of particles of real mirror momentum $\tilde{p}$ having opposite bound state numbers $Q$, and $\overline{Q}=-Q$ should be complex-conjugate to each other. The explicit construction of these eigenvalues was performed in~\cite{Seibold:2022mgg}, and as usual it requires introducing auxiliary S~matrices which account for the non-diagonal part of the S~matrix. A first necessary condition is that the matrix part is such that, regardless of the specific form of the string state, it does not spoil complex conjugation. It is easy to check, in the notation of~\cite{Seibold:2022mgg}, that this is the case: indeed
\begin{equation}
    S^{Qy}\Big(\tilde{x}^\pm,y\Big) =
    \overline{S}{}^{\overline{Q}y}\Big(\frac{1}{\tilde{x}^\mp},y\Big)\,,
\end{equation}
where we used the conjugacy condition~\eqref{eq:mirror_Zh_transormation}. The analogous condition for the dressing factors, which should be seen as a constraint on their (as of yet undetermined) form, is
\begin{equation}
    S^{QQ_a}_{sl}\Big(\tilde{x}^\pm,u_a\Big) = \widetilde{S}^{\overline{Q}Q_a}_{su}\Big(\frac{1}{\tilde{x}^\mp},u_a\Big)\,,
    \qquad
    \widetilde{S}^{Q\overline{Q}_a}_{sl}\Big(\tilde{x}^\pm,u_a\Big) = {S}^{\overline{QQ}_a}_{su}\Big(\frac{1}{\tilde{x}^\mp},u_a\Big)\,,
\end{equation}
again in the notation of~\cite{Seibold:2022mgg}, for real $u_a$ and real mirror momentum.

\section{Production processes in the string and mirror models}
\label{sec:production}

In this section, we prove that $2$-to-$4$ processes are identically null at the tree level both in the string and in the mirror model. This is expected by the fact that the string theory is integrable~\cite{Cagnazzo:2012se} and from a perturbative perspective, integrability reveals itself in the absence of production, as reviewed for some simple examples in~\cite{Dorey:1996gd}. It is however important to check that the absence of production is preserved after the analytic continuation to the mirror model.

We remark that the vanishing of production processes with six external legs is a necessary but not a sufficient condition for the theory to be integrable. In principle, one should prove that all processes with a higher number of external legs ($8$, $10$, $\dots$) vanish too. A general proof could be achieved recursively following the lines of~\cite{Gabai:2018tmm,Bercini:2018ysh, Dorey:2021hub}. However, for the theory under discussion, finding a general proof of absence of production is quite challenging since differently from the Lagrangians considered in~\cite{Gabai:2018tmm,Bercini:2018ysh, Dorey:2021hub} the potential is non-relativistic and contains derivative interactions. It is also worth stressing that the tree-level S-matrices obtained in the previous section are purely elastic and the Yang-Baxter equation is trivially satisfied at the tree level.
The absence of production processes with $6$ external legs is then the first non-trivial check for the integrability of the theory.

\subsection{About factorisation}

Before starting the study of production processes we make some general comments about the factorisation of amplitudes in 1+1 dimensions.
For simplicity, let us consider a standard relativistic propagator
\begin{equation}
\label{eq:relativ_propagator_limit}
    \Delta(p) = \lim_{\epsilon \to 0^+}\,\frac{i}{E^2 - p^2 - m^2+i\epsilon}\, .
\end{equation}
The models under consideration are not relativistic and the propagators have a different form; however, this discussion can be generalised to other types of propagators following the same logic.
Performing the limit in~\eqref{eq:relativ_propagator_limit} we find
\begin{equation}
\begin{split}
\Delta(p) &= \lim_{\epsilon\to 0^+} \frac{i}{E^2-p^2 -m^2 +i\epsilon} \\
&=\lim_{\epsilon\to 0^+}i\,\left( \text{Re}\left(\frac{1}{E^2-p^2 -m^2 +i\epsilon}\right)+i\,\text{Im}\left(\frac{1}{E^2-p^2 -m^2 +i\epsilon}\right)\right)\\
&=\lim_{\epsilon\to 0^+}i\,\left( \frac{E^2-p^2 -m^2}{(E^2-p^2 -m^2)^2 + \epsilon^2} -\frac{i\epsilon}{(E^2 - p^2 -m^2)^2+\epsilon^2}\right)\\
&= i\, \text{p.v.}\Bigl(\frac{1}{E^2 -p^2 -m^2} \Bigl) + \pi \delta(E^2 -p^2 -m^2)\, ,
\end{split}
\end{equation}
where ``p.v.'' denotes the Cauchy principal value.
In order to have a non-vanishing real part, the first term needs to be evaluated outside the pole (\textit{i.e.}, $E^2 -p^2 -m^2 \ne 0$) and corresponds to the propagator when the particle is off-shell; the second term is instead the propagator when the particle is on-shell. This discussion 
holds in the same way for the string and mirror models studied in this paper.  Therefore, for real values of the external energies and momenta, the propagators have poles only in factorised configurations, namely in the configurations in which the number of the incoming particles is equal to the number of the outgoing particles and the set of initial and final momenta are the same. 
In particular, the propagator pole occurs when a particle in the final state has the same momentum as a particle in the initial state. This gives a delta function contribution, which, together with the delta function coming from the conservation of the overall energy-momentum, constrains the scattering to be factorised.

On the other hand, all the processes in which the initial and final sets of momenta are different do not provide any pole to the propagators and the computation can be done in a straightforward way by using the principal value. Therefore, removing $i\epsilon$ from the denominators of propagators corresponds to studying the part of amplitudes associated with production, which we expect to vanish. A more detailed discussion about the factorisation of tree-level amplitudes can be found in~\cite{Dorey:2021hub}.

\subsection{Mapping production between string and mirror model}

To study amplitudes with multiple external particles it is important
to introduce a parameterization for the energies and momenta removing square roots from the dispersion relations. In the string model, at the tree level this can be done through
\begin{equation}
\label{eq:string_energy_momentum_parameterisation}
\begin{split}
&p(m,a)= -m q+\frac{|m|}{2} \sqrt{1-q^2} \bigl(a - \frac{1}{a}\bigl)\,,\\
&\omega(m,a)= \frac{|m|}{2} \sqrt{1-q^2} \bigl(a + \frac{1}{a}\bigl) \,,
\end{split}
\end{equation}
which automatically satisfies the on-shell condition~\eqref{eq:limit_string_energy} for any $a \in \mathbb{C}$. From the positivity condition on the energy, we require that physical particles have $a>0$. Massless particles can be recovered from the limit $\{m \to 0^+,\, a \to 0^+\}$ keeping ${m}/{a}$ fixed, in which case~$p<0$, or from the limit $\{m \to 0^+,\, a \to +\infty \}$ keeping $m\, a$ fixed, in which case~$p>0$. In this manner, we obtain two separate branches for the kinematics of the massless particles.

In a similar fashion, momenta and energies of the mirror particles can be parameterised by
\begin{equation}
\label{eq:mirror_energy_momentum_parameterisation}
\begin{split}
&\tilde{p}(m,a)= \frac{|m|}{2} \sqrt{1-q^2} \bigl(a - \frac{1}{a}\bigl)\,,\\
&\tilde{\omega}(m,a)= i mq+\frac{|m|}{2} \sqrt{1-q^2} \bigl(a + \frac{1}{a}\bigl) \,. 
\end{split}
\end{equation}
Since the mirror particles have imaginary energies, it is non-trivial to establish the values of $a$ characterising the physical region of momenta. Even so, from what we expect in the limit $q \to 0$ and in analogy with the string model, we define the physical region to be $a>0$. Massless particles are identical in the two models and for them the same limiting procedure described above applies.

In the following, we will study tree-level amplitudes with six external legs. After having set $i \epsilon=0$ in the denominators of the propagators, we expect all these amplitudes to be zero: this is due to the fact that all these processes can be mapped to production processes by crossing and the associated amplitudes should be forbidden by integrability. From the parameterizations in~\eqref{eq:string_energy_momentum_parameterisation} and~\eqref{eq:mirror_energy_momentum_parameterisation} we observe that both string and mirror particles satisfy $\omega(m,a)=-\omega(-m,-a)$ and $p(m,a)=-p(-m,-a)$. Due to this fact,
any particle $X(m,a)$ in the initial state is equivalent to an antiparticle $X(-m,-a)$ in the final state and vice-versa.
For this reason, proving the vanishing of a scattering process of the form
\begin{equation}
X(m_1, a_1)+X(m_2, a_2)+ \dots + X(m_N, a_N) \to 0
\end{equation}
is equivalent, for example, to prove the vanishing of
\begin{equation}
X(m_2, a_2)+ \dots + X(m_N, a_N) \to X(-m_1, -a_1) \,.
\end{equation}
It is therefore not restrictive to assume all the particles to be incoming, as incoming particles can be converted into outgoing antiparticles by flipping their quantum numbers $m$. This procedure is known as \textit{crossing}.

It is also interesting to note that the energies and momenta of the string and mirror particles are connected by
\begin{equation}
p(m, i a) = i \tilde{\omega}(m,a) \hspace{3mm} \text{and} \hspace{3mm} \omega(m, i a) = i \tilde{p}(m,a) \,.
\end{equation}
Due to this fact, amplitudes in the string theory can be mapped into amplitudes in the mirror theory by sending $a \to i a$:
\begin{equation}
\mathcal{M}^{(s)} (i a_1, \dots, i a_N) = \mathcal{M}^{(m)} (a_1, \dots, a_N)
\end{equation}
This is true for any number of external legs $N$ and for all types of particles.
Proving the vanishing of a string amplitude $\mathcal{M}^{(s)}$ for all sets of incoming particles parameterised by $a_1, \dots, a_N \in \mathbb{C}$ is then equivalent to proving the vanishing of the associated mirror amplitude $\mathcal{M}^{(m)}$. For this reason, if the vanishing of $\mathcal{M}^{(s)}$ does not depend on the region in which the parameters $a_1, \dots, a_N$ take values and on the value of $q$ then the tree-level integrability of the string theory is mapped into the tree-level integrability of the mirror theory and vice-versa.

In the remaining part of this section, we will show analytically the cancellation of three production amplitudes in the mirror model with six external legs for all values of the parameters $a$ and $q$. 
The cancellation of the remaining six-point production amplitudes has been numerically checked at several random points of the kinematical space.

\subsection{Massive production processes in the mirror model}

We start by studying a scattering process of the form
\begin{equation}
\label{eq:6_to_0_z_process}
\bar{z}(a_1)+\bar{z}(a_2)+\bar{z}(a_3) + z(a_4) + z(a_5) + z(a_6) \to 0 \,,
\end{equation}
which is allowed by the interacting vertices of the theory but should be forbidden by integrability.
As already mentioned the amplitude associated with this process is connected by crossing to the amplitude
\begin{equation}
\label{eq:2_to_4_z_process}
\bar{z}(a_1)+\bar{z}(a_2) \to z(-a_3) + \bar{z}(-a_4) + \bar{z}(-a_5) + \bar{z}(-a_6)
\end{equation}
and the vanishing of~\eqref{eq:6_to_0_z_process} implies the vanishing of~\eqref{eq:2_to_4_z_process}.

The on-shell parameterizations for the energies and momenta of the particles $z$ and $\bar{z}$ are given in~\eqref{eq:mirror_energy_momentum_parameterisation} substituting $m=1$ and $m=-1$ respectively, and the particle propagators are summarized in figure~\ref{image_propagators_string_mirror}. 
\begin{figure}
\begin{center}
    \begin{tikzpicture}
\tikzmath{\y=1.5;}

\draw[] (0.2*\y,0*\y) -- (1.8*\y,0*\y);
\draw[->] (0.8*\y,0.15*\y) -- (1.2*\y,0.15*\y);
\filldraw[black] (0.6*\y,0.3*\y)  node[anchor=west] {\scriptsize{$X(m,p)$}};
\filldraw[black] (2.1*\y,0*\y)  node[anchor=west] {$
= \displaystyle\frac{i}{p_0^2 - (p_1+mq)^2 -m^2(1-q^2) + i\epsilon} $};

\draw[] (0.2*\y,-1*\y) -- (1.8*\y,-1*\y);
\draw[->] (0.8*\y,0.15*\y-1*\y) -- (1.2*\y,0.15*\y-1*\y);
\filldraw[black] (0.6*\y,0.3*\y-1*\y)  node[anchor=west] {\scriptsize{$X(m,p)$}};
\filldraw[black] (2.1*\y,0*\y-1*\y)  node[anchor=west] {$
= \displaystyle\frac{i}{(p_0-im q)^2 - p^2_1 -m^2(1-q^2) + i\epsilon}$};

\end{tikzpicture}
\end{center}
\caption{Propagators of a string (in the top)  and mirror (in the bottom) particle carrying momentum $p$ and a quantum number $m \in \mathbb{Z}$. The arrow indicates the momentum flow. Fundamental particles have $m=0, \, \pm1$.}
\label{image_propagators_string_mirror}
\end{figure}
Summing all the Feynman diagrams composed of two vertices of the type depicted in the first line in figure~\ref{image_vertex_interactions_mirror_theory} (there are 9 diagrams of this type in total contributing to the amplitude) we obtain
\begin{center}
    \begin{tikzpicture}
\tikzmath{\y=1.5;}
\filldraw[black] (-2.2*\y,0.05*\y)  node[anchor=west] {$\mathcal{M}^{(m)}_4=$};
\draw[directed] (-0.5*\y,0*\y) -- (0*\y,0*\y);
\draw[directed] (-0.4*\y,0.4*\y) -- (0*\y,0*\y);
\draw[directed] (-0.4*\y,-0.4*\y) -- (0*\y,0*\y);
\draw[directed] (0*\y,0*\y) -- (1*\y,0*\y);
\draw[directed] (1.5*\y,0*\y) -- (1*\y,0*\y);
\draw[directed] (1.4*\y,0.4*\y) -- (1*\y,0*\y);
\draw[directed] (1.4*\y,-0.4*\y) -- (1*\y,0*\y);
\filldraw[black] (-0.15*\y,0*\y)  node[anchor=west] {\scriptsize{$\bullet$}};
\filldraw[black] (0.85*\y,0*\y)  node[anchor=west] {\scriptsize{$\bullet$}};

\filldraw[black] (0.3*\y,0.15*\y)  node[anchor=west] {\scriptsize{$\bar{z}$}};

\filldraw[black] (-0.5*\y,0.5*\y)  node[anchor=west] {\scriptsize{$\bar{z}(a_1)$}};
\filldraw[black] (-1.2*\y,0*\y)  node[anchor=west] {\scriptsize{$\bar{z}(a_2)$}};
\filldraw[black] (-0.5*\y,-0.5*\y)  node[anchor=west] {\scriptsize{$z(a_4)$}};
\filldraw[black] (1.3*\y,0.5*\y)  node[anchor=west] {\scriptsize{$z(a_5)$}};
\filldraw[black] (1.5*\y,0*\y)  node[anchor=west] {\scriptsize{$z(a_6)$}};
\filldraw[black] (1.3*\y,-0.5*\y)  node[anchor=west] {\scriptsize{$\bar{z}(a_3)$}};

\filldraw[black] (2.2*\y,0*\y)  node[anchor=west] {$+ \ \ \dots \ \ = \ \frac{i}{1-q^2} V_{124}\frac{a_1 a_2 a_4}{(a_1+a_2) (a_2+a4) (a_4+a_1)} V_{356} $};
\filldraw[black] (3.17*\y,-0.5*\y)  node[anchor=west] {$+ \ \ \dots$};

\end{tikzpicture}
\end{center}
where the ellipses in the expression above contain diagrams obtained by permuting in all possible non-equivalent ways the scattered particles. After imposing the overall energy-momentum conservation, which for this process is
\begin{equation}
\label{eq:overall_en_mom_conservation}
\sum_{i=1}^6 a_i=0 \hspace{5mm}, \hspace{5mm} \sum_{i=1}^6 \frac{1}{a_i}=0 \,,
\end{equation}
and having solved the constraints for $a_5$ and $a_6$ in terms of the other kinematical variables, after some tedious computations the amplitude can be expressed in the form
\begin{equation}
   \label{eq:amplitude_production_all_z_4_point}
\begin{aligned}
\mathcal{M}^{(m)}_4= &\;\frac{1}{(8 s_3^2 s_4^2 (s_1 + s_4) (s_3 + s_2 s_4))}\\
&\qquad\times \Bigl[i \sqrt{1-q^2} \Bigl( P_0+q^2 P_2 +q^4 P_4 +q^6 P_6  \Bigl)+ \Bigl( q P_1+q^3 P_3 +q^5 P_5 \Bigl) \Bigl] \,.
\end{aligned}
\end{equation}
The amplitude has been obtained in the gauge $a = 1/2$ and (following the lines of~\cite{Kalousios:2009ey}) has been written in terms of the symmetric polynomials 
\begin{equation}
\begin{aligned}
&s_1 \equiv a_1+a_2+a_3\,, \qquad  &&s_2 \equiv a_1 a_2+a_2 a_3+a_3 a_1\, ,\\
&s_3 \equiv a_1 a_2 a_3\,,\qquad  &&s_4 \equiv a_4 \,.
\end{aligned}
\end{equation}
This is possible by the fact that the three particles carrying kinematical variables $a_1$, $a_2$ and $a_3$ are identical. $P_i$ on the RHS of~\eqref{eq:amplitude_production_all_z_4_point} are polynomials in the variables $s_1$, $s_2$, $s_3$ and $s_4$
whose expressions are listed in appendix~\ref{Appendix_on_production_terms}.

Let us now consider the six-point vertex contributing to this amplitude. In order to find this vertex, we need the expression for the sixth-order Lagrangian. We can find it in the usual way by computing the Lagrangian up to and including $\mathcal{O}(1/T^2)$. Finally, we perform the double Wick rotation to pass from the string to the mirror theory.

In the gauge $a = 1/2$ we obtain
\begin{equation}
\begin{split}
    \mathcal{L}_z^{(6)} &= \frac{1}{4}|z|^4 \left(|z|^2(9|\dot{z}|^2 - |\Acute{z}|^2)-(\dot{z}^2 - \Acute{z}^2)(\do{\Bar{z}}^2 - \Acute{\Bar{z}}^2)\right)\\
    &+ \frac{q}{4}|z|^2\left(|z|^2(\dot{z}\Bar{z}- z\dot{\Bar{z}}) + 2 (\acute{z}^2\Bar{z}\Dot{\Bar{z}} - \Acute{\Bar{z}}^2z\dot{z}) + 6|\dot{z}|^2(z\dot{\Bar{z}} - \Bar{z}\dot{z})\right)\\
    & +\frac{q^2}{4}|\dot{z}|^2\left(-z^2 (\dot{\bar{z}}^2 - \Acute{\Bar{z}}^2) - \Bar{z}^2(\dot{z}^2 - \acute{z}^2)+ 2 |z|^2(|\dot{z}|^2- |\Acute{z}|^2)\right) \, ,
\end{split}
\end{equation}
which generates an additional vertex for the production amplitude.
Inserting the external particles into this vertex we generate an additional contribution to the amplitude which can be schematically represented as
\begin{center}
    \begin{tikzpicture}
\tikzmath{\y=1.5;}
\filldraw[black] (-2.1*\y,0*\y)  node[anchor=west] {$\mathcal{M}^{(m)}_6=$};
\draw[directed] (-0.5*\y,0*\y) -- (0*\y,0*\y);
\draw[directed] (-0.4*\y,0.4*\y) -- (0*\y,0*\y);
\draw[directed] (-0.4*\y,-0.4*\y) -- (0*\y,0*\y);
\draw[directed] (0.5*\y,0*\y) -- (0*\y,0*\y);
\draw[directed] (0.4*\y,0.4*\y) -- (0*\y,0*\y);
\draw[directed] (0.4*\y,-0.4*\y) -- (0*\y,0*\y);
\filldraw[black] (-0.15*\y,0*\y)  node[anchor=west] {\scriptsize{$\bullet$}};
\filldraw[black] (-0.5*\y,0.5*\y)  node[anchor=west] {\scriptsize{$\bar{z}(a_1)$}};
\filldraw[black] (-1.1*\y,0*\y)  node[anchor=west] {\scriptsize{$\bar{z}(a_2)$}};
\filldraw[black] (-0.5*\y,-0.5*\y)  node[anchor=west] {\scriptsize{$z(a_4)$}};
\filldraw[black] (0.3*\y,0.5*\y)  node[anchor=west] {\scriptsize{$z(a_5)$}};
\filldraw[black] (0.5*\y,0*\y)  node[anchor=west] {\scriptsize{$z(a_6)$}};
\filldraw[black] (0.3*\y,-0.5*\y)  node[anchor=west] {\scriptsize{$\bar{z}(a_3)$}};
\end{tikzpicture}
\end{center}
It turns out that this vertex, once the external particles are put on-shell and the constraints~\eqref{eq:overall_en_mom_conservation} are imposed generates exactly the opposite of~\eqref{eq:amplitude_production_all_z_4_point} so that the total tree-level amplitude associated with the process in~\eqref{eq:6_to_0_z_process} vanishes:
\begin{equation}
 \mathcal{M}^{(m)}_{4} + \mathcal{M}^{(m)}_6 = 0\, .
\end{equation}

The computation can similarly be reproduced for $a \ne \frac{1}{2}$; in this case, we obtain additional contributions to the amplitude of order $\bigl(a-\frac{1}{2} \bigl)$ and $\bigl(a-\frac{1}{2} \bigl)^2$. These additional contributions cancel individually and the production amplitude is null for all values of the gauge parameter $a$. We checked this fact numerically at several random points of the kinematical space and several values of $a$.

\subsection{Massless production processes}

Let us now consider production processes in which there are both massless and massive particles. 
As already explained the dispersion relations of massless particles can be recovered from~\eqref{eq:mirror_energy_momentum_parameterisation} by sending simultaneously $m\to0^+$ and $a \to +\infty$ keeping $m a$ fixed or sending $m\to0^+$ and $a \to 0^+$ keeping $\frac{m}{a}$ fixed. The two limits correspond to two different branches of the kinematics of massless particles which we label by
\begin{equation}
\label{eq:massless_chiral_antichiral_branches}
\begin{split}
&\text{Chiral:} \hspace{9mm}  \tilde{p}=\tilde{\omega}_\circ>0 \,,\\
&\text{Antichiral:} \ \  \tilde{p}=-\tilde{\omega}_\circ<0 \,.
\end{split}
\end{equation}

An example of a process involving both massive and massless excitations is
\begin{equation}
\label{eq:massless_process_uuvvzz}
   \bar{u}(k_1) + u(k_2) \to \bar{v}(k_3) + v(k_4) + \bar{z}(a_1) + z(a_2) \,,
\end{equation}
where we labelled the kinematics of the massive particles by the parameters $a_i$, as in~\eqref{eq:mirror_energy_momentum_parameterisation}, and the kinematics of the massless particles by their spatial momenta $k_1$, $k_2$, $k_3$ and $k_4$.
The overall energy-momentum conservation in this mixed notation reads
\begin{equation}
\label{eq:overall_en_mom_mixed_mass}
\begin{aligned}
&k_1 + |k_1|+ k_2 + |k_2| = k_3 + |k_3|+ k_4 + |k_4| + \sqrt{1-q^2}(a_1 + a_2) \, , \\
&|k_1|- k_1 + |k_2| - k_2 = |k_3| - k_3 + |k_4| - k_4 + \sqrt{1-q^2} \bigl(\frac{1}{a_1}+ \frac{1}{a_2} \bigl)\, .
\end{aligned}
\end{equation}
In the following, we solve these constraints for $a_1$ and $a_2$ in terms of $k_1$, $k_2$, $k_3$ and $k_4$.

\begin{figure}
\begin{center}
    \begin{tikzpicture}
\tikzmath{\y=1.5;}
\draw[directed] (-0.5*\y,0*\y) -- (0*\y,0*\y);
\draw[directed] (-0.4*\y,0.4*\y) -- (0*\y,0*\y);
\draw[directed] (0*\y,0*\y) -- (0*\y-0.4*\y,0*\y-0.4*\y);
\draw[directed] (0*\y,0*\y) -- (1*\y,0*\y);
\draw[directed] (1*\y,0*\y) -- (1.5*\y,0*\y);
\draw[directed] (1*\y,0*\y) -- (1.4*\y,0.4*\y);
\draw[directed] (1*\y,0*\y) -- (1.4*\y,-0.4*\y);
\filldraw[black] (-0.15*\y,0*\y)  node[anchor=west] {\scriptsize{$\bullet$}};
\filldraw[black] (0.85*\y,0*\y)  node[anchor=west] {\scriptsize{$\bullet$}};

\filldraw[black] (0.3*\y,0.15*\y)  node[anchor=west] {\scriptsize{$\bar{z}$}};

\filldraw[black] (-0.5*\y,0.5*\y)  node[anchor=west] {\scriptsize{$\bar{u}$}};
\filldraw[black] (-0.8*\y,0*\y)  node[anchor=west] {\scriptsize{$u$}};
\filldraw[black] (-0.5*\y,-0.5*\y)  node[anchor=west] {\scriptsize{$z$}};
\filldraw[black] (1.3*\y,0.5*\y)  node[anchor=west] {\scriptsize{$\bar{v}$}};
\filldraw[black] (1.5*\y,0*\y)  node[anchor=west] {\scriptsize{$v$}};
\filldraw[black] (1.3*\y,-0.5*\y)  node[anchor=west] {\scriptsize{$\bar{z}$}};

\draw[directed] (-0.5*\y+3*\y,0*\y) -- (0*\y+3*\y,0*\y);
\draw[directed] (-0.4*\y+3*\y,0.4*\y) -- (0*\y+3*\y,0*\y);
\draw[directed] (0*\y+3*\y,0*\y) -- (-0.4*\y+3*\y,-0.4*\y);
\draw[directed] (0*\y+3*\y,0*\y) -- (1*\y+3*\y,0*\y);
\draw[directed] (1*\y+3*\y,0*\y) -- (1.5*\y+3*\y,0*\y);
\draw[directed] (1*\y+3*\y,0*\y) -- (1.4*\y+3*\y,0.4*\y);
\draw[directed] (1*\y+3*\y,0*\y) -- (1.4*\y+3*\y,-0.4*\y);
\filldraw[black] (-0.15*\y+3*\y,0*\y)  node[anchor=west] {\scriptsize{$\bullet$}};
\filldraw[black] (0.85*\y+3*\y,0*\y)  node[anchor=west] {\scriptsize{$\bullet$}};

\filldraw[black] (0.3*\y+3*\y,0.15*\y)  node[anchor=west] {\scriptsize{$z$}};

\filldraw[black] (-0.5*\y+3*\y,0.5*\y)  node[anchor=west] {\scriptsize{$\bar{u}$}};
\filldraw[black] (-0.8*\y+3*\y,0*\y)  node[anchor=west] {\scriptsize{$u$}};
\filldraw[black] (-0.5*\y+3*\y,-0.5*\y)  node[anchor=west] {\scriptsize{$\bar{z}$}};
\filldraw[black] (1.3*\y+3*\y,0.5*\y)  node[anchor=west] {\scriptsize{$\bar{v}$}};
\filldraw[black] (1.5*\y+3*\y,0*\y)  node[anchor=west] {\scriptsize{$v$}};
\filldraw[black] (1.3*\y+3*\y,-0.5*\y)  node[anchor=west] {\scriptsize{$z$}};

\draw[directed] (-0.5*\y+6*\y,0*\y) -- (0*\y+6*\y,0*\y);
\draw[directed] (-0.4*\y+6*\y,0.4*\y) -- (0*\y+6*\y,0*\y);
\draw[directed] (0*\y+6*\y,0*\y) -- (-0.4*\y+6*\y,-0.4*\y);
\draw[directed] (0*\y+6*\y,0*\y) -- (1*\y+6*\y,0*\y);
\draw[directed] (1*\y+6*\y,0*\y) -- (1.5*\y+6*\y,0*\y);
\draw[directed] (1*\y+6*\y,0*\y) -- (1.4*\y+6*\y,0.4*\y);
\draw[directed] (1*\y+6*\y,0*\y) -- (1.4*\y+6*\y,-0.4*\y);
\filldraw[black] (-0.15*\y+6*\y,0*\y)  node[anchor=west] {\scriptsize{$\bullet$}};
\filldraw[black] (0.85*\y+6*\y,0*\y)  node[anchor=west] {\scriptsize{$\bullet$}};

\filldraw[black] (0.3*\y+6*\y,0.15*\y)  node[anchor=west] {\scriptsize{$\bar{v}$}};

\filldraw[black] (-0.5*\y+6*\y,0.5*\y)  node[anchor=west] {\scriptsize{$\bar{u}$}};
\filldraw[black] (-0.8*\y+6*\y,0*\y)  node[anchor=west] {\scriptsize{$u$}};
\filldraw[black] (-0.5*\y+6*\y,-0.5*\y)  node[anchor=west] {\scriptsize{$v$}};
\filldraw[black] (1.3*\y+6*\y,0.5*\y)  node[anchor=west] {\scriptsize{$\bar{z}$}};
\filldraw[black] (1.5*\y+6*\y,0*\y)  node[anchor=west] {\scriptsize{$\bar{v}$}};
\filldraw[black] (1.3*\y+6*\y,-0.5*\y)  node[anchor=west] {\scriptsize{$z$}};

\draw[directed] (-0.5*\y,0*\y-2*\y) -- (0*\y,0*\y-2*\y);
\draw[directed] (-0.4*\y,0.4*\y-2*\y) -- (0*\y,0*\y-2*\y);
\draw[directed] (0*\y,0*\y-2*\y) -- (-0.4*\y,-0.4*\y-2*\y);
\draw[directed] (0*\y,0*\y-2*\y) -- (1*\y,0*\y-2*\y);
\draw[directed] (1*\y,0*\y-2*\y) -- (1.5*\y,0*\y-2*\y);
\draw[directed] (1*\y,0*\y-2*\y) -- (1.4*\y,0.4*\y-2*\y);
\draw[directed] (1*\y,0*\y-2*\y) -- (1.4*\y,-0.4*\y-2*\y);
\filldraw[black] (-0.15*\y,0*\y-2*\y)  node[anchor=west] {\scriptsize{$\bullet$}};
\filldraw[black] (0.85*\y,0*\y-2*\y)  node[anchor=west] {\scriptsize{$\bullet$}};

\filldraw[black] (0.3*\y,0.15*\y-2*\y)  node[anchor=west] {\scriptsize{$v$}};

\filldraw[black] (-0.5*\y,0.5*\y-2*\y)  node[anchor=west] {\scriptsize{$\bar{u}$}};
\filldraw[black] (-0.8*\y,0*\y-2*\y)  node[anchor=west] {\scriptsize{$u$}};
\filldraw[black] (-0.5*\y,-0.5*\y-2*\y)  node[anchor=west] {\scriptsize{$\bar{v}$}};
\filldraw[black] (1.3*\y,0.5*\y-2*\y)  node[anchor=west] {\scriptsize{$z$}};
\filldraw[black] (1.5*\y,0*\y-2*\y)  node[anchor=west] {\scriptsize{$v$}};
\filldraw[black] (1.3*\y,-0.5*\y-2*\y)  node[anchor=west] {\scriptsize{$\bar{z}$}};

\draw[directed] (+3*\y,0*\y-2*\y) -- (-0.5*\y+3*\y,0*\y-2*\y);
\draw[directed] (-0.4*\y+3*\y,0.4*\y-2*\y) -- (0*\y+3*\y,0*\y-2*\y);
\draw[directed] (0*\y+3*\y,0*\y-2*\y) -- (-0.4*\y+3*\y,-0.4*\y-2*\y);
\draw[directed] (0*\y+3*\y,0*\y-2*\y) -- (1*\y+3*\y,0*\y-2*\y);
\draw[directed] (1*\y+3*\y,0*\y-2*\y) -- (1.5*\y+3*\y,0*\y-2*\y);
\draw[directed] (1.4*\y+3*\y,0.4*\y-2*\y) -- (1*\y+3*\y,0*\y-2*\y);
\draw[directed] (1*\y+3*\y,0*\y-2*\y) -- (1.4*\y+3*\y,-0.4*\y-2*\y);
\filldraw[black] (-0.15*\y+3*\y,0*\y-2*\y)  node[anchor=west] {\scriptsize{$\bullet$}};
\filldraw[black] (0.85*\y+3*\y,0*\y-2*\y)  node[anchor=west] {\scriptsize{$\bullet$}};

\filldraw[black] (0.3*\y+3*\y,0.15*\y-2*\y)  node[anchor=west] {\scriptsize{$\bar{u}$}};

\filldraw[black] (-0.5*\y+3*\y,0.5*\y-2*\y)  node[anchor=west] {\scriptsize{$\bar{u}$}};
\filldraw[black] (-0.8*\y+3*\y,0*\y-2*\y)  node[anchor=west] {\scriptsize{$v$}};
\filldraw[black] (-0.5*\y+3*\y,-0.5*\y-2*\y)  node[anchor=west] {\scriptsize{$\bar{v}$}};
\filldraw[black] (1.3*\y+3*\y,0.5*\y-2*\y)  node[anchor=west] {\scriptsize{$u$}};
\filldraw[black] (1.5*\y+3*\y,0*\y-2*\y)  node[anchor=west] {\scriptsize{$\bar{z}$}};
\filldraw[black] (1.3*\y+3*\y,-0.5*\y-2*\y)  node[anchor=west] {\scriptsize{$z$}};

\draw[directed] (+6*\y,0*\y-2*\y) -- (-0.5*\y+6*\y,0*\y-2*\y);
\draw[directed] (-0.4*\y+6*\y,0.4*\y-2*\y) -- (0*\y+6*\y,0*\y-2*\y);
\draw[directed] (0*\y+6*\y,0*\y-2*\y) -- (-0.4*\y+6*\y,-0.4*\y-2*\y);
\draw[directed] (0*\y+6*\y,0*\y-2*\y) -- (1*\y+6*\y,0*\y-2*\y);
\draw[directed] (1*\y+6*\y,0*\y-2*\y) -- (1.5*\y+6*\y,0*\y-2*\y);
\draw[directed] (1.4*\y+6*\y,0.4*\y-2*\y) -- (1*\y+6*\y,0*\y-2*\y);
\draw[directed] (1*\y+6*\y,0*\y-2*\y) -- (1.4*\y+6*\y,-0.4*\y-2*\y);
\filldraw[black] (-0.15*\y+6*\y,0*\y-2*\y)  node[anchor=west] {\scriptsize{$\bullet$}};
\filldraw[black] (0.85*\y+6*\y,0*\y-2*\y)  node[anchor=west] {\scriptsize{$\bullet$}};

\filldraw[black] (0.3*\y+6*\y,0.15*\y-2*\y)  node[anchor=west] {\scriptsize{$u$}};

\filldraw[black] (-0.5*\y+6*\y,0.5*\y-2*\y)  node[anchor=west] {\scriptsize{$u$}};
\filldraw[black] (-0.8*\y+6*\y,0*\y-2*\y)  node[anchor=west] {\scriptsize{$v$}};
\filldraw[black] (-0.5*\y+6*\y,-0.5*\y-2*\y)  node[anchor=west] {\scriptsize{$\bar{v}$}};
\filldraw[black] (1.3*\y+6*\y,0.5*\y-2*\y)  node[anchor=west] {\scriptsize{$\bar{u}$}};
\filldraw[black] (1.5*\y+6*\y,0*\y-2*\y)  node[anchor=west] {\scriptsize{$\bar{z}$}};
\filldraw[black] (1.3*\y+6*\y,-0.5*\y-2*\y)  node[anchor=west] {\scriptsize{$z$}};

\end{tikzpicture}
\end{center}
\caption{Feynman diagrams composed of two $4$-point vertices connected by  a propagator contributing to the process in equation~\eqref{eq:massless_process_uuvvzz}.}
\label{image_massive_massless_diagrams_production}
\end{figure}
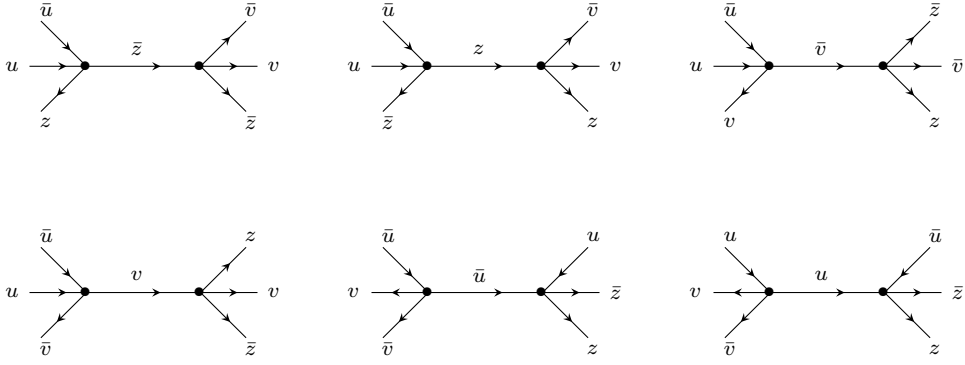

Let us consider all Feynman diagrams composed of only $4$-point vertices (and no $6$-point vertex) first; there are six diagrams of this type contributing to the process in~\eqref{eq:massless_process_uuvvzz}, as depicted in figure~\ref{image_massive_massless_diagrams_production}. 
As before, we label the sum of these diagrams by $\mathcal{M}^{(m)}_4$.

For each massless particle there are two separate branches of the kinematics, as defined in~\eqref{eq:massless_chiral_antichiral_branches}. For this reason, there are in total $16=2^4$ possible configurations for the massless momenta (two for each massless particle involved in the process) and we need to solve the constraints in~\eqref{eq:overall_en_mom_mixed_mass} for all these different configurations. 
Below we focus on the branch $k_1>0>k_2$ for which the incoming massless particles move to opposite directions and the scattering can happen.
Depending on the signs of the momenta of the outgoing massless particles we can have four possible kinematical configurations:
\begin{itemize}
    \item[(1)]  $k_1 > 0$, $k_2 < 0$, $k_3 > 0$, $k_4 < 0$;
    \item[(2)]  $k_1 > 0$, $k_2 < 0$, $k_3 > 0$, $k_4 >0$;
    \item[(3)]  $k_1 > 0$, $k_2 < 0$, $k_3 < 0$, $k_4 >0$;
    \item[(4)]  $k_1 > 0$, $k_2 < 0$, $k_3 < 0$, $k_4 < 0$.
\end{itemize}
The values of the amplitude in these four different kinematical regions are not all independent as the theory has different symmetries. For example, due to the invariance of the mirror Lagrangian under exchanges of $v$  and $\bar{v}$ it follows that the amplitude in configuration (3) can be obtained from the amplitude in configuration (1) by exchanging $k_3$ and $k_4$. 
Similarly configuration (4) is connected to configuration (2) by a parity transformation (switching the signs of the momenta) and exchanging $u$ with $\bar{u}$; these are both symmetries of the mirror Lagrangian. Therefore we report the amplitude only for the kinematical configurations (1) and (2). For configuration~(1) we have,
\begin{equation}
\label{eq:first_mixed_mass_production_amplitude}
\mathcal{M}_4^{(m)}=0 \,,
\end{equation}
while for configuration~(2) we have
\begin{equation}
\label{eq:second_mixed_mass_production_amplitude}
\begin{aligned}
&\mathcal{M}_4^{(m)}= \frac{4  (2 a-1) k_1 k_2 k_3 k_4}{i(k_3+k_4-k_1)} \biggl[\big(k_3+k_4-k_1-(1-2a)  k_2 \big) (1-q^2)\\
&\qquad\qquad\qquad\qquad-2iq (a-1) \sqrt{k_2 (k_3+k_4-k_1)(k_3 k_2+k_4 k_2-k_1 k_2+q^2-1)}\biggl] \,.
\end{aligned}
\end{equation}
Even though, in order to remove the absolute values from the constraints~\eqref{eq:overall_en_mom_mixed_mass}, we set proper signs to the spatial momenta it is important to remark that the expressions in~\eqref{eq:first_mixed_mass_production_amplitude} and~\eqref{eq:second_mixed_mass_production_amplitude} can be analytically continued to all values of the momenta of the massless particles.
For example, considering $k_1<0$ corresponds to transforming the incoming particle $\bar{u}$ into an outgoing particle $u$ with positive energy and momentum.
Ultimately, by solving the Virasoro constraints, it is possible to find the six-point vertex of this process.  Even in this case, we find that this vertex exactly cancels out $\mathcal{M}^{(m)}_4$ and then the total amplitude is zero.

Finally, let us consider a production process involving only massless particles. These processes are all trivially zero if we choose the gauge $a=\frac{1}{2}$; indeed, for $a=\frac{1}{2}$ the vertices involving only massless particles vanish, as we can see from the expressions in figure~\ref{image_vertex_interactions_string_theory_massless}. However, it is possible to check that production processes involving all massless interactions are null in any gauge, as we will show in one moment.

It is important to remark that scattering amplitudes of massless excitations may have `0/0' ambiguities due to vertices and internal propagators vanishing simultaneously. In six-point processes, this can happen in diagrams with two vertices connected by a propagator: when the particles attached to the same vertex are all chiral (or antichiral) then the momentum flowing into the propagator is on-shell and the propagator is singular. However, the residue at the singularity is proportional to a $4$-point amplitude containing massless excitations moving to the same spatial direction. From the expressions in figure~\ref{image_vertex_interactions_string_theory_massless} it is easy to check that these amplitudes are always zero. Therefore, keeping the $i \epsilon$ prescription in the propagators all these ill-defined diagrams vanish from themselves. This is a well-known fact, which is common to several massless theories (see for example~\cite{Hoare:2018jim}).

As an example, let us consider the scattering
\begin{equation}
    u(k_1)+u(k_2)+u(k_3)+\bar{u}(k_4)+\bar{u}(k_5)+\bar{u}(k_6)\to 0\, ,
\end{equation}
whose energy-momentum conservation constraint reads
\begin{equation}
\begin{aligned}
&k_1 + k_2 + k_3 + k_4 + k_5 + k_6 = 0\, ,\\
& |k_1| + |k_2| + |k_3| + |k_4| + |k_5| + |k_6| = 0\, .
\end{aligned}
\end{equation}
Also in this case, in order to remove the absolute values, it is convenient to split the discussion in different cases according to the signs of the momenta. Setting $k_1>0$ and $k_2>0$ we have the following independent kinematical configurations,
\begin{itemize}
    \item $k_3>0$ and $k_4>0$ : 
    \begin{equation}
     \mathcal{M}^{(m)}_4 = 0
    \end{equation}
    \item $k_3<0$ and $k_4>0$:
    \begin{equation}
        \mathcal{M}^{(m)}_4 = 0
    \end{equation}
    \item $k_3<0$ and $k_4<0$ : 
    \begin{equation}
     \mathcal{M}^{(m)}_4 =   -64 i (1-2 a)^2 k_1 k_2 \left(k_1+k_2\right) k_3 k_4 \left(k_3+k_4\right)\, .
    \end{equation}
\end{itemize}
All the other cases can be obtained from these using crossing, parity or time reversal transformations.
As before, we labelled by $\mathcal{M}^{(m)}_4$ the 6-point production amplitude obtained by summing over the Feynman diagrams containing two vertices connected by a propagator.

By expanding the Lagrangian to the next order we obtain
\begin{equation}
\mathcal{L}_u^{(6)} = \frac{1}{2}(1-2a)^2 (\dot{u}^2 - \acute{u}^2) (\dot{u}\dot{\bar{u}} - \acute{u}\acute{\bar{u}}) (\dot{\bar{u}}^2 - \acute{\bar{u}})\, ,
\end{equation}
and it is possible to show that the six-point vertex obtained from this Lagrangian interaction exactly cancels $\mathcal{M}_4$ no matter the kinematical branch considered.
In conclusion, we have seen that all the production processes with six external particles have vanishing amplitudes at the tree level, according to the integrability hypothesis.

\section{Conclusions}

In this paper we have computed the tree-level S~matrix for mixed-flux $AdS_3\times S^3\times T^4$ and for its mirror model. We have found that formally the absence of particle production and the compatibility with the perturbative results hold true, notwithstanding the fact that the mirror dispersion relation $\tilde{\omega}(\tilde{p},m)$ is complex. In fact, the parameter~$q$ can be taken to have any complex value. Therefore, we can in principle think of analytically continuing $q$ to the imaginary axis, which makes the energy real and the perturbative S~matrix well-defined, and analytically continue the computation back to real~$q$.
Of course when computing physical quantities in the original model, such as the L\"uscher corrections to the energy, we must find a well-defined result for~$q$ real. Encouragingly, this is the case. The reason is that for any mirror particle with energy~$\tilde{\omega}(\tilde{p},m)$, there is another with energy 
\begin{equation}
    \tilde{\omega}(\tilde{p},m){}^*= \tilde{\omega}(\tilde{p},-m)\,,\qquad
    \tilde{p}\in\mathbb{R}\,,\quad q\in\mathbb{R}\,.
\end{equation}
Moreover, the S-matrix elements satisfy similar conjugation properties.
These facts are true nonperturbatively, as they descend from the crossing-invariance of the theory~\cite{Lloyd:2014bsa}. As a result, when computing the L\"uscher correction to the energy, the overall contribution is~real. 

In this paper we have only studied the scattering of bosonic states. Dealing with Fermions in the Green-Schwarz formalism is possible but substantially more involved. However, we expect that all excitations of the model arrange themselves in supersymmetric representations of the form described in~\cite{Lloyd:2014bsa} for the string model, or in a suitable analytic continuation of the representations of~\cite{Lloyd:2014bsa} for the mirror model. These representations consist of a quadruplet of two bosons and two fermions with the same energy. As such, our arguments for the reality of the L\"uscher corrections goes through in the complete model (if anything, supersymmetry makes the corrections smaller). It would be of course interesting to check other properties of the model, such as the absence of particle production, for the processes involign fermions too.

While the full, all-loop mirror S-matrix will have a rather complicated analytic structure (see~\cite{Babichenko:2014yaa, Frolov:2023lwd}), our results suggest that we may be able to carry out the construction of the mirror thermodynamic Bethe ansatz notwithstanding the complexity of the energy (and the non-unitarity of the S~matrix) when $q$ is real. Indeed this approach was successfully applied to the derivation of the mirror TBA for the special case~$q=1$~\cite{Dei:2018mfl} (which could be checked against the WZW construction). In that case the S~matrix is particularly simple: it is given just by a CDD factor~\cite{Baggio:2018gct}. Because of that, the TBA construction can be carried out explicitly by interpreting the contributions of $iqm$ as twist of the boundary conditions,
eventually obtaining a real energy for all states.%
\footnote{%
Interestingly, this is also true for the $AdS_3\times S^3\times S^3\times S^1$ mixed-flux background, where the mass spectrum is more complicated~\cite{Dei:2018jyj}.} 
It is worth stressing that only at $q=1$ the imaginary part of the mirror energy can be taken to be a momentum-independent twist \textit{at all loops}; for $0<q<1$ the imaginary part of the energy depends on the momentum when we consider the theory beyond one~loop. Still, it may be possible to live with the lack of reality of the mirror model for real~$q$ without spoiling the properties of physical observables.

It is also worth mentioning that there exists a different approach to compute the spectrum of a string background --- the so-called quantum spectral curve. Originally for $AdS_5\times S^5$~\cite{
Gromov:2013pga}, this was derived by simplifying the TBA equations, but in the case of $AdS_3\times S^3\times T^4$ this has been conjectured based on symmetries and certain analytic assumptions for the $q=0$ case~\cite{Ekhammar:2021pys,Cavaglia:2021eqr}. It is not entirely clear whether such assumptions give the correct spectrum, and how they may be adapted to the case~$0<q<1$. It would be interesting to put that construction on firmer grounds by relating it to the construction of the S~matrix.

\section*{Acknowledgements}
We are grateful to Stijn van Tongeren for bringing the notion of pseudo-unitarity to our attention.
DP and AS thank the participants of the workshop ``Integrability in Low-Supersymmetry Theories'' in Filicudi, Italy, for a stimulating environment where part of this work was carried out.
DP and AS acknowledge support from the European Union - NextGenerationEU, and from the program STARS@UNIPD, under project ``Exact-Holo\-graphy – A new exact approach to holography: harnessing the power of string theory, conformal field theory, and integrable models.''
A.S.\ also acknowledges support from the CARIPLO Foundation ``Supporto ai giovani talenti italiani nelle competizioni dell'European Research Council'' grant n.~2022-1886 ``Nuove basi per la teoria delle stringhe''.

\appendix

\section{Expressions for interaction Lagrangians}
\label{Appendix_on_interacting_Hamiltonians_and_Lagrangians}

In this appendix, we provide the 
interaction Lagrangians of the string and mirror models to the fourth order in the field expansion (which is equivalent to an expansion in large string tension). 

\subsection{Fourth-order Lagrangian of the string model}
\label{appendix:Lagrangian_string_model}
The four-point interaction Lagrangian of the string model can be written in the following compact notation
\begin{equation}
\label{eq:4_point_interacting_Lagrangian_string_model}
\mathcal{L}^{(4)}=\sum_{X=z,y}\mathcal{L}^{(X)} +\mathcal{L}^{(zy)}+\sum_{\mu = u, v}\mathcal{L}^{(\mu)}+ \mathcal{L}^{(uv)}+\sum_{\substack{X=z, y \\ \mu=u, v}}\mathcal{L}^{(X \mu)} \,,
\end{equation}
where the labels $X$ and $\mu$ span the massive and massless fields respectively. Below we summarize the different terms composing~\eqref{eq:4_point_interacting_Lagrangian_string_model}.
Using the definition in~\eqref{eq:sign_X_z_y}, the interactions involving massive fields are
\begin{equation}
\mathcal{L}^{(X)}= \mathscr{S}_X A^{(X)} +(a-\frac{1}{2}) B^{(X)}
\end{equation}
and
\begin{equation}
\mathcal{L}^{(zy)}= C^{(zy)}+(a-\frac{1}{2}) D^{(zy)} \,,
\end{equation}
where
\begin{equation}
\begin{split}
A^{(X)}=& -2|X|^2|\acute{X}|^2
 - \frac{iq}{2} \Bigl( \bigl( |X|^2+|\acute{X}|^2 \bigl) \bigl( X \acute{\bar{X}}-\bar{X} \acute{X} \bigl)+ \dot{X}^2 \bar{X} \acute{\bar{X}} - \dot{\bar{X}}^2 X \acute{X} \Bigl) \,,\\
B^{(X)}=&
 |X|^4- \bigl(\dot{X}^2 - \acute{X}^2 \bigl) \bigl(\dot{\bar{X}}^2 - \acute{\bar{X}}^2 \bigl)\\
-&iq \Bigl( \bigl( X \acute{\bar{X}}-\bar{X} \acute{X} \bigl)  \bigl( |\acute{X}|^2-|X|^2 \bigl)+ \dot{X}^2 \bar{X} \acute{\bar{X}} - \dot{\bar{X}}^2 X \acute{X}  \Bigl)  \,,
\end{split}
\end{equation}
\begin{equation}
\begin{split}
C^{(zy)}&= |y|^2 \bigl(|\dot{z}|^2+|\acute{z}|^2 \bigl)-|z|^2 \bigl( |\dot{y}|^2+|\acute{y}|^2 \bigl)
-i\frac{q}{2} \Bigl[ \bigl(|z|^2-|\acute{z}|^2-|\dot{z}|^2 \bigl) \bigl(y \acute{\bar{y}}- \bar{y} \acute{y} \bigl)\\
&-\bigl( |y|^2-|\acute{y}|^2-|\dot{y}|^2 \bigl) \bigl(  z \acute{\bar{z}}-\bar{z} \acute{z} \bigl) +\bigl( \acute{z} \dot{\bar{z}}+\acute{\bar{z}} \dot{z} \bigl) \bigl( y \dot{\bar{y}}- \bar{y} \dot{y} \bigl) - \bigl(\acute{y} \dot{\bar{y}}+\acute{\bar{y}}\dot{y} \bigl) \bigl( z \dot{\bar{z}}-\bar{z} \dot{z} \bigl) \Bigl] \,, \\
D^{(zy)}&= 2 |z|^2|y|^2+  2 \bigl( \acute{z} \dot{\bar{z}}+\acute{\bar{z}} \dot{z} \bigl) \bigl(\acute{y} \dot{\bar{y}}+\acute{\bar{y}} \dot{y} \bigl) - 2 \bigl( |\dot{z}|^2+|\acute{z}|^2 \bigl) \bigl(|\dot{y}|^2+|\acute{y}|^2 \bigl)\\
&-iq\Bigl[ \bigl(z \acute{\bar{z}}-\bar{z} \acute{z} \bigl)\bigl(|\dot{y}|^2+|\acute{y}|^2-|y|^2 \bigl)+\bigl( y \acute{\bar{y}}-\bar{y} \acute{y} \bigl)  \bigl(|\dot{z}|^2+|\acute{z}|^2-|z|^2\bigl)\\
&\qquad\qquad+\bigl(\acute{z} \dot{\bar{z}}+\acute{\bar{z}} \dot{z} \bigl) \bigl(\bar{y} \dot{y} - y \dot{\bar{y}} \bigl)+\bigl( \acute{y} \dot{\bar{y}}+\acute{\bar{y}} \dot{y} \bigl) \bigl(\bar{z} \dot{z}-z \dot{\bar{z}} \bigl)\Bigl] \,.
\end{split}
\end{equation}

The interactions involving massless fields are instead
\begin{equation}
\begin{split}
\mathcal{L}^{(\mu)}&=\Bigl(a-\frac{1}{2} \Bigl) \Bigl(  \bigl(\acute{\mu} \dot{\bar{\mu}}+\acute{\bar{\mu}} \dot{\mu} \bigl)^2 - \bigl(|\dot{\mu}|^2 +|\acute{\mu}|^2 \bigl)^2 \Bigl) \quad \mu = u, v \,,\\
\mathcal{L}^{(uv)}&=(2a-1)\Bigl( (\acute{u} \dot{\bar{u}}+\acute{\bar{u}} \dot{u})(\acute{v} \dot{\bar{v}}+\acute{\bar{v}} \dot{v})-(|\dot{u}|^2 +|\acute{u}|^2)(|\dot{v}|^2+|\acute{v}|^2) \Bigl) \,.
\end{split}
\end{equation}
We remark that these interactions disappear in the gauge $a=\frac{1}{2}$. 

Finally, the interactions mixing massive and massless fields are
\begin{equation}
\mathcal{L}^{(X\mu)}= \mathscr{S}_X E^{(X\mu)}+(2a-1)F^{(X\mu)}
\end{equation}
with
\begin{equation}
\begin{split}
E^{(X\mu)} =&-|X|^2 (|\dot{\mu}|^2+|\acute{\mu}|^2)\\
&-i\frac{q}{2} \Bigl( \bigl( |\dot{\mu}|^2+|\acute{\mu}|^2 \bigl) \bigl( X \acute{\bar{X}} -\bar{X} \acute{X} \bigl) -  \bigl( \acute{\mu} \dot{\bar{\mu}}+\acute{\bar{\mu}} \dot{\mu} \bigl) \bigl( X \dot{\bar{X}} -\bar{X} \dot{X} \bigl)\Bigl) \,,\\
F^{(X\mu)} =& \bigl( \acute{X} \dot{\bar{X}}+\acute{\bar{X}} \dot{X} \bigl)\bigl( \acute{\mu} \dot{\bar{\mu}}+\acute{\bar{\mu}} \dot{\mu} \bigl)-\bigl( |\dot{X}|^2 +|\acute{X}|^2 \bigl) \bigl( |\dot{\mu}|^2+|\acute{\mu}|^2 \bigl) \\
&-\frac{iq}{2} \Bigl(  \bigl( \acute{\mu} \dot{\bar{\mu}}+\acute{\bar{\mu}} \dot{\mu} \bigl) \bigl( \bar{X} \dot{X}-X \dot{\bar{X}} \bigl) -\bigl( \bar{X} \acute{X} - X \acute{\bar{X}} \bigl)  \bigl( |\dot{\mu}|^2+|\acute{\mu}|^2 \bigl)  \Bigl)  \, .
\end{split}
\end{equation}

\subsection{Fourth-order Lagrangian of the mirror model}
\label{appendix:Lagrangian_mirror_model}

Applying a double Wick rotation, as described in section~\ref{sec:mirror_theory}, it is possible to derive the quartic Lagrangian of the mirror model starting from the Lagrangian of the string model. As for the string model, this Lagrangian can be as
\begin{equation}
\label{eq:4_point_interacting_Lagrangian_mirror_model}
\mathcal{L}_m^{(4)}=\sum_{X=z,y}\mathcal{L}_m^{(X)} +\mathcal{L}_m^{(zy)}+\sum_{\mu = u, v}\mathcal{L}_m^{(\mu)}+ \mathcal{L}_m^{(uv)}+\sum_{\substack{X=z, y \\ \mu=u, v}}\mathcal{L}_m^{(X \mu)} \,.
\end{equation}
The massive interactions contributing to~\eqref{eq:4_point_interacting_Lagrangian_mirror_model} are given by
\begin{equation}
\mathcal{L}_m^{(X)}= \mathscr{S}_XA_m^{(X)}+(2a-1) B_m^{(X)}
\end{equation}
and
\begin{equation}
\mathcal{L}_m^{(zy)}= C_m^{(zy)}+(2a-1) D_m^{(zy)} \,,
\end{equation}
with
\begin{equation}
\begin{split}
A_m^{(X)}=& +2|X|^2|\dot{X}|^2
 - \frac{q}{2} \Bigl( \bigl( |X|^2-|\dot{X}|^2 \bigl) \bigl( X \dot{\bar{X}}-\bar{X} \dot{X} \bigl)- \acute{X}^2 \bar{X} \dot{\bar{X}} + \acute{\bar{X}}^2 X \dot{X} \Bigl) \,,\\
B_m^{(X)}=&
 |X|^4- \bigl(\acute{X}^2 - \dot{X}^2 \bigl) \bigl(\acute{\bar{X}}^2 - \dot{\bar{X}}^2 \bigl)\\
-&q \Bigl( - \bigl( X \dot{\bar{X}}-\bar{X} \dot{X} \bigl)  \bigl( |\dot{X}|^2+|X|^2 \bigl)- \acute{X}^2 \bar{X} \dot{\bar{X}} + \acute{\bar{X}}^2 X \dot{X}  \Bigl)  \,,
\end{split}
\end{equation}
\begin{equation}
\begin{split}
C_m^{(zy)}=& -|y|^2 \bigl(|\dot{z}|^2+|\acute{z}|^2 \bigl)+|z|^2 \bigl( |\dot{y}|^2+|\acute{y}|^2 \bigl)
-\frac{q}{2} \Bigl[ \bigl(|z|^2+|\dot{z}|^2+|\acute{z}|^2 \bigl) \bigl(y \dot{\bar{y}}- \bar{y} \dot{y} \bigl)\\
&-\bigl( |y|^2+|\dot{y}|^2 +|\acute{y}|^2\bigl) \bigl(  z \dot{\bar{z}}-\bar{z} \dot{z} \bigl) -\bigl( \acute{z} \dot{\bar{z}}+\acute{\bar{z}} \dot{z} \bigl) \bigl( y \acute{\bar{y}}- \bar{y} \acute{y} \bigl) + \bigl(\acute{y} \dot{\bar{y}}+\acute{\bar{y}}\dot{y} \bigl) \bigl( z \acute{\bar{z}}-\bar{z} \acute{z} \bigl) \Bigl] \,,\\
D_m^{(zy)}=& = 2 |z|^2|y|^2+  2 \bigl( \acute{z} \dot{\bar{z}}+\acute{\bar{z}} \dot{z} \bigl) \bigl(\acute{y} \dot{\bar{y}}+\acute{\bar{y}} \dot{y} \bigl) - 2 \bigl( |\dot{z}|^2+|\acute{z}|^2 \bigl) \bigl(|\dot{y}|^2+|\acute{y}|^2 \bigl)\\
&-q\Bigl[ \bigl(\bar{z} \dot{z} -z \dot{\bar{z}}\bigl)\bigl(|\dot{y}|^2+|\acute{y}|^2+|y|^2 \bigl)
+\bigl(\bar{y} \dot{y}- y \dot{\bar{y}} \bigl)  \bigl(|\dot{z}|^2+|\acute{z}|^2+|z|^2\bigl)\\
&\qquad\qquad+\bigl(\acute{z} \dot{\bar{z}}+\acute{\bar{z}} \dot{z} \bigl)  \bigl(y \acute{\bar{y}}-\bar{y} \acute{y} \bigl)+  \bigl( \acute{y} \dot{\bar{y}}+\acute{\bar{y}} \dot{y} \bigl) \bigl(z \acute{\bar{z}}-\bar{z} \acute{z} \bigl)\Bigl] \,.
\end{split}
\end{equation}
As defined in~\eqref{eq:sign_X_z_y} $\mathscr{S}_X=1$ if $X=z$ and $\mathscr{S}_X=-1$ if $X=y$.

The massless interactions are
\begin{equation}
\begin{split}
\mathcal{L}_m^{(\mu)}&=\Bigl(a-\frac{1}{2} \Bigl) \Bigl(  \bigl(\acute{\mu} \dot{\bar{\mu}}+\acute{\bar{\mu}} \dot{\mu} \bigl)^2 - \bigl(|\dot{\mu}|^2 +|\acute{\mu}|^2 \bigl)^2 \Bigl) \quad \mu = u, v \,,\\
\mathcal{L}_m^{(uv)}&=(2a-1)\Bigl( (\acute{u} \dot{\bar{u}}+\acute{\bar{u}} \dot{u})(\acute{v} \dot{\bar{v}}+\acute{\bar{v}} \dot{v})-(|\dot{u}|^2 +|\acute{u}|^2)(|\dot{v}|^2+|\acute{v}|^2) \Bigl) \,,
\end{split}
\end{equation}
and are invariant under the mirror transformation.

The interactions between massive and massless fields are
\begin{equation}
\mathcal{L}^{(X\mu)}= \mathscr{S}_X E_m^{(X\mu)}+(2a-1)F_m^{(X\mu)}
\end{equation}
with
\begin{equation}
\begin{split}
E_m^{(X\mu)} =&|X|^2 (|\acute{\mu}|^2+|\dot{\mu}|^2)\\
&+\frac{q}{2} \Bigl( \bigl( |\acute{\mu}|^2+|\dot{\mu}|^2 \bigl) \bigl( X \dot{\bar{X}} -\bar{X} \dot{X} \bigl) -  \bigl( \acute{\mu} \dot{\bar{\mu}}+\acute{\bar{\mu}} \dot{\mu} \bigl) \bigl( X \acute{\bar{X}} -\bar{X} \acute{X} \bigl)\Bigl) \,,\\
F_m^{(X\mu)} =& \bigl( \acute{X} \dot{\bar{X}}+\acute{\bar{X}} \dot{X}  \bigl) \bigl( \acute{\mu} \dot{\bar{\mu}}+\acute{\bar{\mu}} \dot{\mu} \bigl)-\bigl( |\dot{X}|^2 +|\acute{X}|^2 \bigl) \bigl( |\dot{\mu}|^2+|\acute{\mu}|^2 \bigl) \\
&-\frac{q}{2} \Bigl( \bigl(\bar{X} \dot{X} - X \dot{\bar{X}} \bigl)  \bigl(|\dot{\mu}|^2+|\acute{\mu}|^2 \bigl)  - \bigl(\bar{X} \acute{X}-X \acute{\bar{X}} \bigl) \bigl(\acute{\mu} \dot{\bar{\mu}}+\acute{\bar{\mu}} \dot{\mu} \bigl) \Bigl) \, .
\end{split}
\end{equation}

\section{Amplitude production terms}
\label{Appendix_on_production_terms}

In this appendix, we list the different terms composing the production amplitude in~\eqref{eq:amplitude_production_all_z_4_point}.
\begin{equation}
P_0= 2 (8 s_2^2 - s_1^2 s_2^2 + 26 s_1 s_2 s_3 - 9 s_3^2 + 8 s_1^2 s_3^2) s_4^2 (s_1 +
    s_4) (s_3 + s_2 s_4) \,,
\end{equation}
\begin{equation}
\begin{split}
&P_1=-4 s_1 s_2 s_3^2 - 4 s_1^2 s_3^3 - 8 s_1 s_2^2 s_3 s_4 - 8 s_1^2 s_2 s_3^2 s_4 \\
&- 4 s_1 s_2^3 s_4^2 + 4 s_1^2 s_2 s_3 s_4^2 + 4 s_1^2 s_2^2 s_3 s_4^2 + 
 4 s_1^3 s_3^2 s_4^2 + 168 s_1 s_2 s_3^2 s_4^2 + 8 s_1^3 s_2 s_3^2 s_4^2 + 
 4 s_1 s_2^2 s_3^2 s_4^2 \\
 &+ 168 s_1^2 s_3^3 s_4^2 + 4 s_1^2 s_2 s_3^3 s_4^2 + 
 4 s_1^2 s_2^2 s_4^3 + 8 s_1^2 s_2^3 s_4^3 + 4 s_1 s_2 s_3 s_4^3 + 
 4 s_1^3 s_2 s_3 s_4^3 + 176 s_1 s_2^2 s_3 s_4^3 \\
 &+ 8 s_1^3 s_2^2 s_3 s_4^3+ 4 s_1 s_2^3 s_3 s_4^3 + 4 s_1^2 s_3^2 s_4^3 + 168 s_2 s_3^2 s_4^3 + 
 176 s_1^2 s_2 s_3^2 s_4^3 + 4 s_2^2 s_3^2 s_4^3 + 4 s_1^2 s_2^2 s_3^2 s_4^3 \\
 &+ 
 168 s_1 s_3^3 s_4^3 + 4 s_1 s_2 s_3^3 s_4^3 + 4 s_1 s_2^2 s_4^4 + 
 8 s_1 s_2^3 s_4^4 + 4 s_1^2 s_2 s_3 s_4^4 + 168 s_2^2 s_3 s_4^4 \\
 &+ 4 s_1^2 s_2^2 s_3 s_4^4 + 4 s_2^3 s_3 s_4^4 + 168 s_1 s_2 s_3^2 s_4^4 - 4 s_1^3 s_2 s_3^2 s_4^4 \\
 &+ 4 s_1 s_2^2 s_3^2 s_4^4 - 8 s_1 s_2^2 s_3 s_4^5 - 
 8 s_1^2 s_2 s_3^2 s_4^5 - 4 s_2^2 s_3 s_4^6 - 4 s_1 s_2 s_3^2 s_4^6 \,,
 \end{split}
\end{equation}
\begin{equation}
\begin{split}
&P_2= 3 s_1^2 s_2 s_3^2 + 21 s_1 s_3^3 + 2 s_1^3 s_3^3 + 2 s_1 s_2 s_3^3 + 
 6 s_1^2 s_2^2 s_3 s_4 + 42 s_1 s_2 s_3^2 s_4 + 4 s_1^3 s_2 s_3^2 s_4 \\
 &+ 4 s_1 s_2^2 s_3^2 s_4 + 3 s_1^2 s_2^3 s_4^2 - 3 s_1^3 s_2 s_3 s_4^2 - 
 27 s_1 s_2^2 s_3 s_4^2 + 5 s_1^3 s_2^2 s_3 s_4^2 - 21 s_1^2 s_3^2 s_4^2 - 
 2 s_1^4 s_3^2 s_4^2\\
 &- 194 s_1^2 s_2 s_3^2 s_4^2 - 3 s_1^2 s_2^2 s_3^2 s_4^2 - 
 261 s_1 s_3^3 s_4^2 - 48 s_1^3 s_3^3 s_4^2 - 21 s_1 s_2 s_3^3 s_4^2 - 
 3 s_1^3 s_2^2 s_4^3 - 48 s_1 s_2^3 s_4^3\\
 &+ 3 s_1^3 s_2^3 s_4^3 - 
 2 s_1 s_2^4 s_4^3- 24 s_1^2 s_2 s_3 s_4^3 - 2 s_1^4 s_2 s_3 s_4^3 - 
 48 s_2^2 s_3 s_4^3 - 191 s_1^2 s_2^2 s_3 s_4^3 - 2 s_2^3 s_3 s_4^3 \\
 &- 3 s_1^2 s_2^3 s_3 s_4^3 - 21 s_1 s_3^2 s_4^3 - 2 s_1^3 s_3^2 s_4^3 - 
 455 s_1 s_2 s_3^2 s_4^3 - 48 s_1^3 s_2 s_3^2 s_4^3 - 24 s_1 s_2^2 s_3^2 s_4^3 -  261 s_3^3 s_4^3\\
 &- 48 s_1^2 s_3^3 s_4^3 - 21 s_2 s_3^3 s_4^3 - 
 3 s_1^2 s_2^2 s_4^4 - 48 s_2^3 s_4^4 + 5 s_1^2 s_2^3 s_4^4 - 2 s_2^4 s_4^4- 21 s_1 s_2 s_3 s_4^4 - 194 s_1 s_2^2 s_3 s_4^4\\
 & + 3 s_1^3 s_2^2 s_3 s_4^4- 3 s_1 s_2^3 s_3 s_4^4 - 261 s_2 s_3^2 s_4^4 - 27 s_1^2 s_2 s_3^2 s_4^4 - 21 s_2^2 s_3^2 s_4^4 + 4 s_1 s_2^3 s_4^5 + 4 s_1^2 s_2 s_3 s_4^5 \\
 &+ 6 s_1^2 s_2^2 s_3 s_4^5 + 42 s_1 s_2 s_3^2 s_4^5 + 2 s_2^3 s_4^6 + 
 2 s_1 s_2 s_3 s_4^6 + 3 s_1 s_2^2 s_3 s_4^6 + 21 s_2 s_3^2 s_4^6 \,,
 \end{split}
\end{equation}
\begin{equation}
\begin{split}
&P_3=8 s_1 s_2 s_3^2 + 24 s_1^2 s_3^3 + 16 s_1 s_2^2 s_3 s_4 + 48 s_1^2 s_2 s_3^2 s_4 + 8 s_1 s_2^3 s_4^2 - 8 s_1^2 s_2 s_3 s_4^2 + 
 8 s_1 s_2 s_3^2 s_4^6\\
&- 24 s_1^3 s_3^2 s_4^2 - 
 264 s_1 s_2 s_3^2 s_4^2 - 24 s_1^3 s_2 s_3^2 s_4^2 - 24 s_1 s_2^2 s_3^2 s_4^2 - 264 s_1^2 s_3^3 s_4^2 - 8 s_1^2 s_2 s_3^3 s_4^2 \\
 &- 8 s_1^2 s_2^2 s_4^3 - 
 24 s_1^2 s_2^3 s_4^3 - 8 s_1 s_2 s_3 s_4^3 - 24 s_1^3 s_2 s_3 s_4^3 - 
 288 s_1 s_2^2 s_3 s_4^3 - 24 s_1^3 s_2^2 s_3 s_4^3 \\
 &- 24 s_1 s_2^3 s_3 s_4^3 - 
 24 s_1^2 s_3^2 s_4^3 - 264 s_2 s_3^2 s_4^3 - 288 s_1^2 s_2 s_3^2 s_4^3 - 
 24 s_2^2 s_3^2 s_4^3 - 8 s_1^2 s_2^2 s_3^2 s_4^3 \\
 &- 264 s_1 s_3^3 s_4^3 - 
 8 s_1 s_2 s_3^3 s_4^3 - 8 s_1 s_2^2 s_4^4 - 24 s_1 s_2^3 s_4^4 - 
 24 s_1^2 s_2 s_3 s_4^4- 264 s_2^2 s_3 s_4^4 - 24 s_2^3 s_3 s_4^4 \\
 &- 
 264 s_1 s_2 s_3^2 s_4^4 + 8 s_1^3 s_2 s_3^2 s_4^4 - 8 s_1 s_2^2 s_3^2 s_4^4 + 
 48 s_1 s_2^2 s_3 s_4^5 + 16 s_1^2 s_2 s_3^2 s_4^5 + 24 s_2^2 s_3 s_4^6 \,,
 \end{split}
\end{equation}
\begin{equation}
\begin{split}
&P_4=-6 s_1^2 s_2 s_3^2 - 42 s_1 s_3^3 - 4 s_1^3 s_3^3 - 4 s_1 s_2 s_3^3 - 
 12 s_1^2 s_2^2 s_3 s_4 - 84 s_1 s_2 s_3^2 s_4 - 8 s_1^3 s_2 s_3^2 s_4\\
 &- 6 s_1^2 s_2^3 s_4^2 + 6 s_1^3 s_2 s_3 s_4^2+ 6 s_1 s_2^2 s_3 s_4^2 - 4 s_1^3 s_2^2 s_3 s_4^2 + 42 s_1^2 s_3^2 s_4^2 + 4 s_1^4 s_3^2 s_4^2 + 256 s_1^2 s_2 s_3^2 s_4^2 \\
 &+ 6 s_1^2 s_2^2 s_3^2 s_4^2 + 
 360 s_1 s_3^3 s_4^2 + 48 s_1^3 s_3^3 s_4^2 + 42 s_1 s_2 s_3^3 s_4^2 + 
 6 s_1^3 s_2^2 s_4^3 + 48 s_1 s_2^3 s_4^3 + 4 s_1 s_2^4 s_4^3\\
 &+ 48 s_1^2 s_2 s_3 s_4^3 + 4 s_1^4 s_2 s_3 s_4^3 + 48 s_2^2 s_3 s_4^3 + 
 256 s_1^2 s_2^2 s_3 s_4^3 + 4 s_2^3 s_3 s_4^3 + 6 s_1^2 s_2^3 s_3 s_4^3 + 42 s_1 s_3^2 s_4^3\\
 &+ 4 s_1^3 s_3^2 s_4^3 + 616 s_1 s_2 s_3^2 s_4^3 + 48 s_1^3 s_2 s_3^2 s_4^3 + 48 s_1 s_2^2 s_3^2 s_4^3 + 360 s_3^3 s_4^3 + 48 s_1^2 s_3^3 s_4^3 + 42 s_2 s_3^3 s_4^3 \\
 &+ 6 s_1^2 s_2^2 s_4^4 + 
 48 s_2^3 s_4^4 - 4 s_1^2 s_2^3 s_4^4 + 4 s_2^4 s_4^4 + 42 s_1 s_2 s_3 s_4^4 + 256 s_1 s_2^2 s_3 s_4^4- 6 s_1^3 s_2^2 s_3 s_4^4 \\
 &+ 6 s_1 s_2^3 s_3 s_4^4 + 360 s_2 s_3^2 s_4^4 + 6 s_1^2 s_2 s_3^2 s_4^4 + 42 s_2^2 s_3^2 s_4^4 - 
 8 s_1 s_2^3 s_4^5 - 
 8 s_1 s_2^2 s_3^2 s_4\\
 &- 8 s_1^2 s_2 s_3 s_4^5 - 12 s_1^2 s_2^2 s_3 s_4^5 - 84 s_1 s_2 s_3^2 s_4^5 - 4 s_2^3 s_4^6 - 4 s_1 s_2 s_3 s_4^6 - 
 6 s_1 s_2^2 s_3 s_4^6 - 42 s_2 s_3^2 s_4^6 \,,
 \end{split}
\end{equation}
\begin{equation}
\begin{split}
&P_5=-4 s_1 s_2 s_3^2 - 20 s_1^2 s_3^3 - 8 s_1 s_2^2 s_3 s_4 - 40 s_1^2 s_2 s_3^2 s_4 - 
 4 s_1 s_2^3 s_4^2 + 4 s_1^2 s_2 s_3 s_4^2 \\
 &- 4 s_1^2 s_2^2 s_3 s_4^2 + 
 20 s_1^3 s_3^2 s_4^2 + 96 s_1 s_2 s_3^2 s_4^2 + 16 s_1^3 s_2 s_3^2 s_4^2 + 20 s_1 s_2^2 s_3^2 s_4^2 + 96 s_1^2 s_3^3 s_4^2 + 4 s_1^2 s_2 s_3^3 s_4^2 \\
 &+ 
 4 s_1^2 s_2^2 s_4^3 + 16 s_1^2 s_2^3 s_4^3 + 4 s_1 s_2 s_3 s_4^3 + 
 20 s_1^3 s_2 s_3 s_4^3 + 112 s_1 s_2^2 s_3 s_4^3 + 16 s_1^3 s_2^2 s_3 s_4^3 + 
 20 s_1 s_2^3 s_3 s_4^3 \\
 &+ 20 s_1^2 s_3^2 s_4^3 + 96 s_2 s_3^2 s_4^3 + 
 112 s_1^2 s_2 s_3^2 s_4^3 + 20 s_2^2 s_3^2 s_4^3 + 4 s_1^2 s_2^2 s_3^2 s_4^3 + 
 96 s_1 s_3^3 s_4^3 \\
 &+ 4 s_1 s_2 s_3^3 s_4^3 + 4 s_1 s_2^2 s_4^4 + 
 16 s_1 s_2^3 s_4^4 + 20 s_1^2 s_2 s_3 s_4^4 + 96 s_2^2 s_3 s_4^4- 
 4 s_1^2 s_2^2 s_3 s_4^4 + 20 s_2^3 s_3 s_4^4 \\
 &+ 96 s_1 s_2 s_3^2 s_4^4 - 
 4 s_1^3 s_2 s_3^2 s_4^4 + 4 s_1 s_2^2 s_3^2 s_4^4 - 40 s_1 s_2^2 s_3 s_4^5 - 
 8 s_1^2 s_2 s_3^2 s_4^5 - 20 s_2^2 s_3 s_4^6 - 4 s_1 s_2 s_3^2 s_4^6 \,.
 \end{split}
\end{equation}

\begin{equation}
\begin{split}
&P_6=3 s_1^2 s_2 s_3^2 + 21 s_1 s_3^3 + 2 s_1^3 s_3^3 + 2 s_1 s_2 s_3^3 + 
 6 s_1^2 s_2^2 s_3 s_4 + 42 s_1 s_2 s_3^2 s_4 + 4 s1^3 s2 s3^2 s4 \\
 &+ 4 s_1 s_2^2 s_3^2 s_4 + 3 s_1^2 s_2^3 s_4^2 - 3 s_1^3 s_2 s_3 s_4^2 + 
 5 s_1 s_2^2 s_3 s_4^2 + s_1^3 s_2^2 s_3 s_4^2 - 21 s_1^2 s_3^2 s_4^2 \\
 &- 2 s_1^4 s_3^2 s_4^2 - 114 s_1^2 s_2 s_3^2 s_4^2 - 3 s_1^2 s_2^2 s_3^2 s_4^2 - 81 s_1 s_3^3 s_4^2 - 16 s_1^3 s_3^3 s_4^2 - 21 s_1 s_2 s_3^3 s_4^2 + 21 s_2 s_3^2 s_4^6\\
 &- 3 s_1^3 s_2^2 s_4^3 - 16 s_1 s_2^3 s_4^3 - s_1^3 s_2^3 s_4^3 - 
 2 s_1 s_2^4 s_4^3 - 24 s_1^2 s_2 s_3 s_4^3 - 2 s_1^4 s_2 s_3 s_4^3 - 
 16 s_2^2 s_3 s_4^3\\
 &- 115 s_1^2 s_2^2 s_3 s_4^3 - 2 s_2^3 s_3 s_4^3 - 
 3 s_1^2 s_2^3 s_3 s_4^3 - 21 s_1 s_3^2 s_4^3 - 2 s_1^3 s_3^2 s_4^3 - 
 195 s_1 s_2 s_3^2 s_4^3 - 16 s_1^3 s_2 s_3^2 s_4^3 \\
 &- 24 s_1 s_2^2 s_3^2 s_4^3 - 
 81 s_3^3 s_4^3 - 16 s_1^2 s_3^3 s_4^3 - 21 s_2 s_3^3 s_4^3 - 
 3 s_1^2 s_2^2 s_4^4 - 16 s_2^3 s_4^4 + s_1^2 s_2^3 s_4^4 - 2 s_2^4 s_4^4\\
 &- 21 s_1 s_2 s_3 s_4^4 - 114 s_1 s_2^2 s_3 s_4^4 + 3 s_1^3 s_2^2 s_3 s_4^4 - 3 s_1 s_2^3 s_3 s_4^4 - 81 s_2 s_3^2 s_4^4 + 5 s_1^2 s_2 s_3^2 s_4^4 - 21 s_2^2 s_3^2 s_4^4 \\
 &+ 4 s_1 s_2^3 s_4^5 + 4 s_1^2 s_2 s_3 s_4^5 +6 s_1^2 s_2^2 s_3 s_4^5 + 42 s_1 s_2 s_3^2 s_4^5 + 2 s_2^3 s_4^6 + 
 2 s_1 s_2 s_3 s_4^6 + 3 s_1 s_2^2 s_3 s_4^6
\end{split}
\end{equation}

\bibliographystyle{JHEP}
\bibliography{refs}

\end{document}